\documentclass[aps,pre,twocolumn,showpacs,floatfix,superscriptaddress]{revtex4-1}
\usepackage{graphicx,amsfonts,amssymb,amsmath,hyperref}
\usepackage{textcomp}
\usepackage{esint}

\newif\ifhyper
\hypertrue
\ifhyper       
\hypersetup{        
  citecolor = {green},
  colorlinks = {true}, 
  urlcolor = {blue} 
}

\begin{document}

\def\Tkt{T_{\rm KT}}
\def\aopt{\alpha_{\rm opt}}
\def\trhor{\trho_{\rm r}}


\graphicspath{{./figures_submit/}}
\def\rhoeq{\hat\rho_{\rm eq}}

\newcommand{\marge}[1]{\marginpar{\scriptsize #1}}
\newcommand{\remarque}[1]{\marginpar{\scriptsize Remarque}{\it [#1]}}
\newcommand{\new}[1]{{\bf #1}}
\newcommand{\red}[1]{\textcolor{red}{#1}}
\newcommand{\blue}[1]{\textcolor{blue}{#1}}
\newlength{\textlarg}
\newcommand{\barre}[1]{%
   \settowidth{\textlarg}{#1}
   #1\hspace{-\textlarg}\rule[0.5ex]{\textlarg}{0.5pt}}
\newcommand{\barred}[1]{%
   \settowidth{\textlarg}{#1}
   \red{#1}\hspace{-\textlarg}\rule[0.5ex]{\textlarg}{0.5pt}}
\newcommand{\barblue}[1]{%
   \settowidth{\textlarg}{#1}
   \blue{#1}\hspace{-\textlarg}\rule[0.5ex]{\textlarg}{0.5pt}}

\def\beq{\begin{equation}}
\def\eeq{\end{equation}}
\def\bleq{\begin{eqnarray}}
\def\eleq{\end{eqnarray}} 
\def\bfig{\begin{figure}}
\def\efig{\end{figure}}
\def\bline{\begin{multline}}
\def\eline{\end{multline}}
\def\bremark{\begin{quotation} \noindent \small }
\def\eremark{\end{quotation}}
\def\llbrace{\left\lbrace}
\def\rrbrace{\right\rbrace}
\def\lbraket{\left[}
\def\rbraket{\right]}
\def\llangle{\left\langle}
\def\rrangle{\right\rangle} 

\newcommand{\Tr}{{\rm Tr}} 
\newcommand{\tr}{{\rm tr}} 
\newcommand{\sgn}{{\rm sgn}} 
\newcommand{\mean}[1]{\langle #1 \rangle}
\newcommand{\commu}[2]{[#1,#2]} 
\newcommand{\bra}[1]{\langle#1|}
\newcommand{\ket}[1]{|#1\rangle}
\newcommand{\braket}[2]{\langle #1|#2\rangle}
\newcommand{\dbraket}[3]{\langle #1|#2|#3\rangle}
\newcommand{\tens}[1]{\overleftrightarrow{#1}}  
\newcommand{\vac}{|{\rm vac}\rangle} 
\def\bravac{\langle{\rm vac}|}
\newcommand{\const}{{\rm const}} 
\newcommand{\atanh}{\,{\rm atanh}}

\newcommand{\ie}{i.e. }
\newcommand{\iet}{i.e.}
\newcommand{\eg}{e.g. }
\newcommand{\cc}{{\rm c.c.}} 
\newcommand{\hc}{{\rm h.c.}} 
\def\etal{{\it et al. }}

\newcommand{\jhatbf}{\hat {\textbf \j}} 
\newcommand{\Jhatbf}{\hat {\textbf \J}} 
\newcommand{\jhat}{\hat {\jmath}} 
\newcommand{\Jhat}{\hat {J}} 
\newcommand{\jbf}{\textbf j}
\newcommand{\Jbf}{\textbf J}

\def\chibf{\boldsymbol{\chi}}
\def\down{\downarrow}
\def\eps{\epsilon}
\def\gam{\gamma} 
\def\phibf{\boldsymbol{\phi}}
\def\varphibf{\boldsymbol{\varphi}}
\def\varphibfs{\boldsymbol{\varphi}_<}
\def\varphibfl{\boldsymbol{\varphi}_>}
\def\varphis{\varphi_{<}}
\def\varphil{\varphi_{>}}
\def\psibf{\boldsymbol{\psi}}
\def\Ome{\Omega}
\def\omeD{{\omega_D}} 
\def\bfOme{\boldsymbol{\Omega}} 
\def\Omebf{\boldsymbol{\Omega}} 
\def\lamb{\lambda}
\def\Lamb{\Lambda}
\def\sig{\sigma}
\def\Sig{\Sigma}
\def\sigp{{\sigma'}} 
\def\bfsig{\boldsymbol{\sigma}} 
\def\sigbf{\boldsymbol{\sigma}} 
\def\The{\Theta} 
\def\up{\uparrow}

\def\epsk{\epsilon_{\bf k}} 
\def\xik{\xi_{\bf k}} 
\def\txik{\tilde\xi_{\bf k}} 
\def\xip{\xi_{\bf p}} 
\def\xikq{\xi_{{\bf k}+{\bf q}}} 
\def\Ek{E_{\bf k}} 
\def\Ep{E_{\bf p}}
\def\Heff{\hat H_{\rm eff}}
\def\Hem{\hat H_{\rm em}}
\def\Hint{\hat H_{\rm int}}
\def\Hloc{\hat H_{\rm loc}}
\def\HMF{\hat H_{\rm MF}}
\def\Sem{S_{\rm em}}
\def\SMF{S_{\rm MF}} 
\def\SHF{S_{\rm HF}} 
\def\SRPA{S_{\rm RPA}} 
\def\Sint{S_{\rm int}} 
\def\Sloc{S_{\rm loc}}
\def\TN{T_{\rm N}} 
\def\TNHF{T^{\rm HF}_{\rm N}} 
\def\Zloc{Z_{\rm loc}} 
\def\ZMF{Z_{\rm MF}} 
\def\ZHF{Z_{\rm HF}} 
\def\ZRPA{Z_{\rm RPA}} 
\def\RPA{{\rm RPA}}
\def\loc{{\rm loc}} 
\def\pp{{\rm pp}}
\def\ph{{\rm ph}} 
\def\ch{{\rm ch}}
\def\sp{{\rm sp}} 
\def\qtf{q_{\rm TF}}
\def\epstf{\eps^{}_{\rm TF}} 
\def\epsrpa{\eps^{}_{\rm RPA}} 
\def\chinnzpp{\chi_{nn}^{0}{}\!\!\!''}

\def\half{\frac{1}{2}}
\def\dhalf{\dfrac{1}{2}}
\def\third{\frac{1}{3}} 
\def\quarter{\frac{1}{4}}

\def\qr{{\bf q}\cdot{\bf r}}
\def\wt{\omega t} 

\def\a{{\bf a}}
\def\b{{\bf b}}
\def\e{{\bf e}}
\def\f{{\bf f}}
\def\g{{\bf g}}
\def\h{{\bf h}}
\def\k{{\bf k}}
\def\l{{\bf l}}
\def\m{{\bf m}}
\def\n{{\bf n}} 
\def\p{{\bf p}} 
\def\q{{\bf q}}
\def\r{{\bf r}}
\def\t{{\bf t}}
\def\u{{\bf u}}
\def\v{{\bf v}}
\def\x{{\bf x}}
\def\y{{\bf y}} 
\def\z{{\bf z}} 
\def\A{{\bf A}}
\def\B{{\bf B}}
\def\D{{\bf D}} 
\def\E{{\bf E}} 
\def\F{{\bf F}} 
\def\H{{\bf H}}  
\def\J{{\bf J}}
\def\K{{\bf K}} 

\def\G{{\bf G}}
\def\L{{\bf L}}
\def\M{{\bf M}}  
\def\O{{\bf O}} 
\def\P{{\bf P}} 
\def\Q{{\bf Q}} 
\def\R{{\bf R}}
\def\S{{\bf S}}
\def\epsbf{\boldsymbol{\epsilon}}
\def\mubf{\boldsymbol{\mu}}
\def\nablabf{\boldsymbol{\nabla}}
\def\rhobf{\boldsymbol{\rho}}
\def\sigmabf{\boldsymbol{\sigma}} 
\def\Pibf{\boldsymbol{\Pi}}
\def\pibf{\boldsymbol{\pi}}

\def\para{\parallel}
\def\kpara{{k_\parallel}}
\def\kperp{{k_\perp}} 
\def\kperpp{{k_\perp'}} 
\def\qperp{{q_\perp}} 
\def\tperp{{t_\perp}} 

\def\w{\omega}
\def\wn{\omega_n}
\def\wnu{\omega_\nu}
\def\wp{\omega_p} 
\def\dmu{{\partial_\mu}}
\def\dl{{\partial_l}}  
\def\dt{\partial_t} 
\def\tdt{\tilde\partial_t}
\def\dk{\partial_k}
\def\tdk{\tilde\partial_k}
\def\dx{\partial_x}
\def\dy{\partial_y} 
\def\dtau{{\partial_\tau}}  
\def\det{{\rm det}} 
\def\Pf{{\rm Pf}}

\def\dsum{\displaystyle \sum}
\def\dint{\displaystyle \int} 
\def\intt{\int_{-\infty}^\infty dt} 
\def\inttp{\int_{-\infty}^\infty dt'} 
\def\intk{\int_{\bf k}} 
\def\intkd{\int \frac{d^dk}{(2\pi)^d}}
\def\intq{\int_{\bf q}} 
\def\intr{\int d^dr}  
\def\dintr{\displaystyle \int d^dr} 
\def\intrp{\int d^dr'}
\def\dinttau{\displaystyle \int_0^\beta d\tau}
\def\dinttaup{\displaystyle \int_0^\beta d\tau'}
\def\inttau{\int_0^\beta d\tau}
\def\inttaup{\int_0^\beta d\tau'}
\def\intx{\int d^{d+1}x} 
\def\inttaur{\int_0^\beta d\tau \int d^dr}
\def\intinf{\int_{-\infty}^\infty}
\def\dinttaur{\displaystyle \int_0^\beta d\tau \int d^dr}
\def\dintinf{\displaystyle \int_{-\infty}^\infty}
\def\intw{\int_{-\infty}^\infty \frac{d\w}{2\pi}}
\def\sumr{\sum_{\bf r}} 

\def\calA{{\cal A}} 
\def\calC{{\cal C}} 
\def\dt{\partial_t}
\def\calD{{\cal D}}
\def\calF{{\cal F}} 
\def\calG{{\cal G}}
\def\calH{{\cal H}}
\def\calI{{\cal I}}
\def\calJ{{\cal J}}
\def\calK{{\cal K}}
\def\calL{{\cal L}} 
\def\calN{{\cal N}}
\def\calO{{\cal O}}
\def\calP{{\cal P}}  
\def\calR{{\cal R}} 
\def\calS{{\cal S}}
\def\calT{{\cal T}}
\def\calU{{\cal U}}
\def\calX{{\cal X}} 
\def\calY{{\cal Y}} 
\def\calZ{{\cal Z}} 

\def\calFbf{{\bf F}}

\def\tT{{\tilde T}}
\def\talpha{{\tilde\alpha}}
\def\tdelta{{\tilde\delta}}
\def\teta{{\tilde\eta}} 
\def\tlamb{{\tilde\lambda}}
\def\tmu{{\tilde\mu}}
\def\tphibf{{\tilde\phibf}}
\def\trho{{\tilde\rho}}
\def\tvarphibf{{\tilde\varphibf}} 
\def\tw{{\tilde\omega}}
\def\twn{{\tilde\omega_n}}

\def\asinh{{\rm asinh}} 


\title{Reexamination of the nonperturbative renormalization-group
approach to the Kosterlitz-Thouless transition} 

\author{P. Jakubczyk}
\affiliation{Institute of Theoretical Physics, Faculty of Physics, University of Warsaw, Ho\.za 69, 00-681 Warsaw, Poland}

\author{N. Dupuis}
\author{B. Delamotte}
\affiliation{Laboratoire de Physique Th\'eorique de la Mati\`ere Condens\'ee, 
CNRS UMR 7600, Universit\'e Pierre et Marie Curie, 4 Place Jussieu, 
75252 Paris Cedex 05, France} 

\date{November 24, 2014} 

\begin{abstract}
We reexamine the two-dimensional linear O(2) model ($\varphi^4$ theory) in the framework of the nonperturbative renormalization-group. From the flow equations obtained in the derivative expansion to second order and with optimization of the infrared regulator, we find a transition between a high-temperature (disordered) phase and a low-temperature phase displaying a line of fixed points and algebraic order. We obtain a picture in agreement with the standard theory of the Kosterlitz-Thouless (KT) transition and reproduce the universal features of the transition. In particular, we find the anomalous dimension $\eta(\Tkt)\simeq 0.24$ and the stiffness jump $\rho_s(\Tkt^-)\simeq 0.64$ at the transition temperature $\Tkt$, in very good agreement with the exact results $\eta(\Tkt)=1/4$ and $\rho_s(\Tkt^-)=2/\pi$, as well as an essential singularity of the 
correlation length in the high-temperature phase as $T\to \Tkt$. 
\end{abstract}
\pacs{05.10.Cc,05.70.Fh,74.20.-z}
\maketitle

\section{Introduction}

The Kosterlitz-Thouless (KT) transition occurs in two-dimensional systems with global O(2) symmetry such as the two-dimensional XY model~\cite{Berezinskii70,*Berezinskii71,Kosterlitz73,Kosterlitz74}. It has been observed in liquid helium films~\cite{Rudnick78,Bishop78,Maps81,Maps82}, array of Josephson junctions~\cite{Resnick81}, trapped two-dimensional atomic gases~\cite{Hadzibabic06,Clade09,Tung10,Desbuquois12}, etc. 

The KT transition differs from more conventional finite-temperature phase transitions in a number of aspects. It is not characterized by spontaneous symmetry breaking and the low-temperature phase exhibits algebraic order (rather than true long-range order).  
Nevertheless, the system shows a nonzero ``stiffness'' $\rho_s(T)$ for all temperatures $T<\Tkt$. Above the transition temperature $\Tkt$, one observes a standard disordered phase with exponentially decaying correlation functions. However, the correlation length $\xi$ does not diverge as a power law of $\tau=T-\Tkt$ but shows an essential singularity $\xi\sim \exp(c/\sqrt{\tau})$. The transition is also characterized by a jump of the stiffness which vanishes for $T>\Tkt$ and takes the universal value $2/\pi$ for $T\to \Tkt^-$~\cite{Nelson77a,Minnhagen81}.  

The key role of topological defects (vortices) was recognized by Kosterlitz and Thouless who formulated the KT transition as a vortex/anti-vortex pair unbinding transition~\cite{Kosterlitz73,Kosterlitz74,Jose77,Ambegaokar80,Minnhagen87}. Standard studies of the KT transition explicitly introduce the vortices in the analysis and use a mapping to the Coulomb gas or sine-Gordon models. A perturbative renormalization-group approach is then sufficient to derive the universal features of the KT transition. 

The KT transition in the two-dimensional linear O(2) model ($\varphibf^4$ theory for a two-component vector field) provides an important benchmark for the nonperturbative renormalization group (NPRG). 
A distinctive feature of the NPRG approach is that the vortices are not introduced explicitly~\cite{Graeter95,Gersdorff01,[{The NPRG has also been used to study the KT transition in the two-dimensional sine-Gordon model: see }] Nagy09} and thus the RG equations are the standard ones of the $d$-dimensional O($N$) model with $N=2$ and $d=2$. In the approach of Gersdorff and Wetterich (GW)~\cite{Gersdorff01}, the KT transition is not captured {\it stricto sensu} since the correlation length is always finite. Nevertheless, below a ``transition'' temperature 
$\Tkt$ one finds a line of quasi-fixed points implying a very large correlation length (although not infinite as expected in the low-temperature phase of the KT transition). Furthermore, the essential scaling of the correlation length $\xi$ above $\Tkt$ is reproduced except in the immediate vicinity of $\Tkt$. Thus, although the NPRG approach by GW does not yield a low-temperature phase with an infinite correlation 
length, it nevertheless allows one to estimate the KT transition temperature and reproduce most of the universal features of the transition. 

Using a lattice version of the NPRG, $\Tkt$ has been computed with reasonable accuracy
for the ferromagnetic XY model on the square lattice~\cite{Machado10}. The NPRG approach
has also been used to study two-dimensional superconductors~\cite{Krahl07} and bosonic superfluids~\cite{Floerchinger09a,Rancon12b,Rancon13b,Rancon14a}. The superfluid transition temperature in a two-dimensional Bose gas, 
with or without an optical lattice~\cite{Rancon12b,Rancon13b}, deduced from the NPRG approach turns out to be in very good agreement with Monte Carlo 
simulations~\cite{Prokofev01,Prokofev02,Capogrosso08}.  

In spite of these successes, the NPRG approach to the two-dimensional linear O(2) model is not fully satisfying. First, from a conceptual point of view, one would like to find a true transition between a high-temperature phase with exponentially decaying correlations and a low-temperature phase exhibiting algebraic order and a line of fixed points. 
Second, from a more practical point of view, we expect the NPRG approach to yield reasonable estimates not only of the transition temperature $\Tkt$ and the anomalous dimension $\eta(\Tkt)$ but also of the temperature dependence of the anomalous dimension $\eta(T)$ and the stiffness $\rho_s(T)$ (including the value of $\rho_s(\Tkt^-)$) in the low-temperature phase, which has not been possible so far due to the absence of a line of true fixed points at low temperatures.

In this paper, we reconsider the NPRG approach to the two-dimensional linear O(2) model. While our RG equations are the same as those of GW~\cite{Gersdorff01}, we explore various ways to set up the RG procedure. In particular we use the freedom in the choice of the infrared regulator and the way the anomalous dimension is computed. The commonly used exponential regulator~\cite{Wetterich93a} with an arbitrary prefactor $\alpha$ considered as a variational parameter, along with a fixed renormalization point (see Sec.~\ref{subsec_num2D}), allows us to find a transition with all expected features of the KT transition. In the high-temperature phase, we reproduce the essential singularity of the 
correlation length as $T\to\Tkt$. In the low-temperature phase, for all $T\leq \Tkt$, it is possible to find a value of the variational parameter $\alpha$ such that the RG flow is attracted by a (true) fixed point of the RG equations. The resulting line of fixed points characterizes a phase with no spontaneous symmetry breaking, 
algebraic order (i.e. $\xi=\infty$), and 
nonzero anomalous dimension $\eta(T)$ and stiffness $\rho_s(T)$. At the transition, we find $\eta(\Tkt)\simeq 0.24$ and $\rho_s(\Tkt^-)\simeq 0.64$, in very good agreement with the exact results $\eta(\Tkt)=1/4$ and $\rho_s(\Tkt^-)=2/\pi\simeq 0.6366$. 

\section{NPRG approach}
\label{sec_nprg} 

The linear O(2) model is defined by the action 
\begin{equation}
S[\varphibf] = \int d^dr \biggl\lbrace \half (\nablabf\varphibf)^2 
+ \frac{r_0}{2} \varphibf^2 + \frac{u_0}{4!} {(\varphibf^2)}^2 \biggr\rbrace, 
\label{action1} 
\end{equation}
where $\varphibf=(\varphi_1,\varphi_2)$ is a two-component real field. For the sake of generality we consider an arbitrary dimension $d$. The model is regularized by an ultraviolet momentum cutoff $\Lambda$. In practice, we consider $u_0$ as a fixed parameter and vary $r_0\propto T-T_0$ to explore the phases of the system ($T_0$ denotes the mean-field transition temperature). 

The strategy of the NPRG approach is to build a family of theories indexed by a momentum scale $k$ such that fluctuations are smoothly taken into account as $k$ is lowered from the microscopic scale $\Lambda$ down to 0~\cite{Berges02,Delamotte12,Kopietz_book}. This is achieved by adding to the action (\ref{action1}) the infrared regulator
\begin{equation}
\Delta S_k[\varphibf] = \half \int \frac{d^dp}{(2\pi)^d} \sum_i \varphi_i(-\p) R_k(\p) \varphi_i(\p) ,
\end{equation}
so that the partition function
\begin{equation}
Z_k[\J] = \int\calD[\varphibf]\, e^{-S[\varphibf]-\Delta S_k[\varphibf] + \int d^dr \J\cdot \varphibf } 
\end{equation}
becomes $k$ dependent. The scale-dependent effective action 
\begin{equation}
\Gamma_k[\phibf] = - \ln Z_k[\J] + \int d^dr  \J \cdot \phibf - \Delta S_k[\phibf] 
\end{equation}
is defined as a modified Legendre transform of $-\ln Z_k[\J]$ which includes the subtraction of $\Delta S_k[\phibf]$. Here $\phibf(\r)=\mean{\varphibf(\r)}$ is the order parameter (in the presence of the external source $\J$). 

The initial condition of the flow is specified by the microscopic scale $k=\Lambda$ where fluctuations are frozen by the $\Delta S_k$ term, so that $\Gamma_\Lambda[\phibf]=S[\phibf]$. The effective action of the original model (\ref{action1}) is given by $\Gamma_{k=0}$ provided that $R_{k=0}$ vanishes. For a generic value of $k$, 
the regulator $R_k(\p)$ suppresses fluctuations with momentum $|\p|\lesssim k$ but leaves unaffected those with $|\p| \gtrsim k$. We use an exponential regulator
\begin{equation}
R_k(\p) = Z_k \p^2 r (\p^2/k^2), \qquad r(y) = \frac{\alpha}{e^y-1} ,
\label{Rdef}
\end{equation}
with an arbitrary parameter $\alpha>0$. The $k$-dependent constant $Z_k$ is defined below. 

The variation of the effective action with $k$ is given by Wetterich's equation~\cite{Wetterich93} 
\begin{equation}
\dt \Gamma_k[\phibf] = \half \Tr\llbrace \dt R_k\left(\Gamma^{(2)}_k[\phibf] + R_k\right)^{-1} \rrbrace ,
\label{rgeq}
\end{equation}
where $t=\ln(k/\Lambda)$. $\Gamma^{(2)}_k[\phibf]$ denotes the second functional derivative of $\Gamma_k[\phibf]$. In Fourier space, the trace involves a sum over momenta as well as the internal index $i=1,2$ of the $\phibf$ field.

To solve the RG equation~(\ref{rgeq}), we use a derivative expansion of $\Gamma_k[\phibf]$~\cite{Berges02,Delamotte12}. To second order, 
\begin{equation}
\Gamma_k[\phibf] = \int d^dr \llbrace \half Z_k(\rho) (\nablabf\phibf)^2 + \quarter Y_k(\rho) (\nablabf\rho)^2 + U_k(\rho) \rrbrace ,
\label{gamde} 
\end{equation}
where $U_k(\rho)$, $Z_k(\rho)$ and $Y_k(\rho)$ are functions of the O(2) invariant $\rho=\phibf^2/2$. For a uniform field, the effective action $\Gamma_k[\phibf]=VU_k(\rho)$ reduces to the effective potential $U_k(\rho)$ ($V$ denotes the system volume). There are two $\calO(\nablabf^2)$ terms, reflecting the fact that longitudinal and transverse fluctuations (wrt the local order parameter $\phibf(\r)$) are characterized by different stiffnesses. Within the ansatz~(\ref{gamde}), the flow equation~(\ref{rgeq}) reduces to three coupled partial differential equations for the functions $U_k(\rho)$, $Z_k(\rho)$ and $Y_k(\rho)$. Since in two dimensions the engineering dimension of the field is zero it is {\it a priori} important
to keep the full field-dependence of these functions and not approximate them by a finite-order field expansion.
Notice that even if the initial condition is polynomial the flow generates all terms allowed by symmetries (up to
second order in derivatives within our approach).

To solve numerically the flow equations and look for possible fixed points, it is convenient to introduce dimensionless and renormalized quantities,
\begin{equation}
\begin{gathered}
\tilde U_k(\trho) = v_d^{-1} k^{-d} U_k(\rho), \quad \tilde Z_k(\trho) = Z_k^{-1} Z_k(\rho), \\ \tilde Y_k(\trho) = v_d Z_k^{-2} k^{d-2} Y_k(\rho) ,
\end{gathered}  
\label{dimless}
\end{equation}
where $\trho=v_d^{-1} Z_kk^{2-d}\rho$. The factor $v_d^{-1}=2^{d+1} \pi^{d/2} \Gamma(d/2)$ is introduced for convenience (see Appendix~\ref{app_rgeq}). The $k$-dependent constant $Z_k$ is defined by imposing the condition $\tilde Z_k(\trhor)=1$ where $\trhor$ is an arbitrary renormalization point. The (running) anomalous dimension is then defined by 
\begin{equation}
\eta_k = - k\dk \ln Z_k , 
\label{eta}
\end{equation}
while the (true) anomalous dimension is simply $\eta=\lim_{k\to 0} \eta_k$. The flow equations are given in Appendix~\ref{app_rgeq}. 

The effective potential $U_k(\rho)$, and the location $\rho_{0,k}$ of its minimum (corresponding to the equilibrium state), provide information about the phase of the system. In a disordered phase with a finite correlation length $\xi$, $\rho_{0,k}$ vanishes for a value of $k$ of the order of $\xi^{-1}$. A nonzero value of $\lim_{k\to 0}\rho_{0,k}$ would imply spontaneous symmetry breaking and is forbidden by the Mermin-Wagner theorem when $d=2$. It is however possible that $\rho_{0,k}$ vanishes as a power law of $k$, which is in fact the expected result when the order is algebraic ($\xi=\infty$). 

Additional information can be obtained from the longitudinal and transverse parts of the propagator in a uniform field~\cite{note2},  
\begin{align}
G_{k,\rm L}(\p;\rho) &= [(Z_k(\rho)+\rho Y_k(\rho)) \p^2 + U_k'(\rho) + 2\rho U_k''(\rho) ]^{-1} , \nonumber \\
G_{k,\rm T}(\p;\rho) &= [Z_k(\rho) \p^2 + U_k'(\rho) ]^{-1} .
\label{GLT} 
\end{align} 
In the equilibrium field configuration and for $\rho_{0,k}>0$, the running stiffness $\rho_{s,k}$ is defined by writing the transverse propagator as
\begin{equation}
G_{k,\rm T}(\p;\rho_{0,k}) = \frac{2\rho_{0,k}}{\rho_{s,k} \p^2} ,
\label{GT}
\end{equation}
where $2\rho_{0,k}=\mean{\varphibf(\r)}^2$ is the square of the order parameter at scale $k$. Note that the momentum dependence of $G_{k,\rm T}$ in Eq.~(\ref{GT}) follows from the derivative expansion and is therefore valid only for $|\p|\ll k$.
Alternatively, $\rho_{s,k}$ can be defined from the change $\Delta\Gamma_k$ of the effective action when the direction of the order parameter $\phibf(\r)=\sqrt{2\rho_{0,k}}(\cos\theta(\r),\sin\theta(\r))$ at scale $k$ varies slowly in space,
\begin{equation}
\Delta\Gamma_k[\phibf] = \half \rho_{s,k} \int d^dr \, (\nablabf\theta)^2 . 
\label{dgam}
\end{equation}
Equations~(\ref{GT}) and (\ref{dgam}) lead to the same expression
\begin{equation}
\rho_{s,k} = 2 Z_k(\rho_{0,k}) \rho_{0,k}
\label{stiff} 
\end{equation} 
of the stiffness. The physical stiffness is defined as $\rho_s(T)=\lim_{k\to 0}\rho_{s,k}(T)$.

Equations~(\ref{eta}) and (\ref{stiff}) are crucial to understand the long-distance behavior of the system when $d=2$. In the high-temperature phase both $\rho_{s,k}$ and $\rho_{0,k}$ vanish for a nonzero value of $k$ of the order of the inverse of the correlation length $\xi$. $Z_k$ reaches a finite limit for $k\to 0$ since the anomalous dimension $\eta=\lim_{k\to 0}\eta_k$ vanishes. In the low-temperature phase, we expect $\eta_k$ and $\rho_{s,k}$ to take a finite value in the limit $k\to 0$ (this implies $Z_k\sim k^{-\eta}$ for $k\to 0$). This is possible only if $\rho_{0,k}\sim k^\eta$ when $k\to 0$, which is consistent with $\trho_{0,k}=Z_k\rho_{0,k}$ taking a finite limit (as expected for a critical system). The result $\rho_{0,k}\sim k^\eta$ is in agreement with both the absence of long-range order ($\lim_{k\to 0}\rho_{0,k}=0$) and an infinite correlation length ($\rho_{0,k}>0$ for any $k>0$).

\section{Numerical integration of RG equations} 
\label{sec_numerics}

\subsection{Optimized regulator for $d=3$} 
\label{subsec_num3D}

\begin{table}
\renewcommand{\arraystretch}{1.25}
\caption{Critical exponents $\nu$ and $\eta$ in the three-dimensional O(2) model obtained from the derivative expansion to second order and the principle of minimum sensitivity. Also shown are the results obtained from field theory (FT) and Monte Carlo (MC) simulations.}
\label{table_N2} 
\begin{tabular}{cccc}
\hline\hline
 & NPRG & FT~\cite{Pogorelov08} & MC~\cite{Campostrini06}  
\\ \hline 
$\nu$ & 0.6707  & 0.6700(6) &  0.6717(1)
\\ \hline 
$\eta$ & 0.047 & 0.0334(2) & 0.0381(2)
\\ \hline \hline
\end{tabular} 
\caption{Same as table~\ref{table_N2} but for the O(3) model.}
\label{table_N3} 
\begin{tabular}{cccc}
\hline\hline
 & NPRG & FT~\cite{Pogorelov08} & MC~\cite{Campostrini02}
\\ \hline 
$\nu$ & 0.719 & 0.7060(7)  & 0.7112(5)
\\ \hline 
$\eta$ & 0.0463 & 0.0333(3)  & 0.0375(5)
\\ \hline \hline
\end{tabular} 
\end{table}

Let us first briefly review the determination of the critical exponents in three dimensions. One can either integrate the flow equations for various initial conditions until a fixed point is reached, or linearize the flow equations about the fixed-point solution defined by $\dt\tilde U^*=\dt\tilde Z^*=\dt\tilde Y^*=0$. If the flow equation of the effective action $\Gamma_k[\phibf]$ were solved exactly, the results would be independent of the infrared regulator $R_k$. This is not the case when the effective action is expanded to second order in a derivative expansion. In particular, with the regulator~(\ref{Rdef}), the critical exponents depend on the parameter $\alpha$. 
We determine what we consider as the optimal value of $\alpha$ from the principle of minimal sensitivity (PMS)~\cite{Canet03b}, i.e. by demanding that locally the critical exponents are independent of $\alpha$ (e.g. for the correlation-length exponent $\nu$, $d\nu/d\alpha=0$ for $\alpha=\aopt$). 
The renormalization point $\tilde\rho_r$ is taken fixed (for numerical convenience) and, provided the fixed point exists, a change in $\trhor$ is equivalent to a change in $\alpha$~\cite{note6} so that the critical exponents obtained from the PMS are independent of $\trhor$. Thus the $k$-dependent renormalization point $\trho_{0,k}$, which becomes $k$ independent at small $k$ since $\trho_{0,k}\to\trho_0^*$ at criticality, is equivalent to any other choice $\trhor=\const$.  

The results for the critical exponents $\nu$ and $\eta$ are shown in Tables~\ref{table_N2} and \ref{table_N3} for the three-dimensional O(2) and O(3) models. They compare very well with results from field theory (resummed perturbative theory) and Monte Carlo simulations. 

\subsection{Optimized regulator for $d=2$}
\label{subsec_num2D}

\begin{figure}
\centerline{\includegraphics[width=7cm]{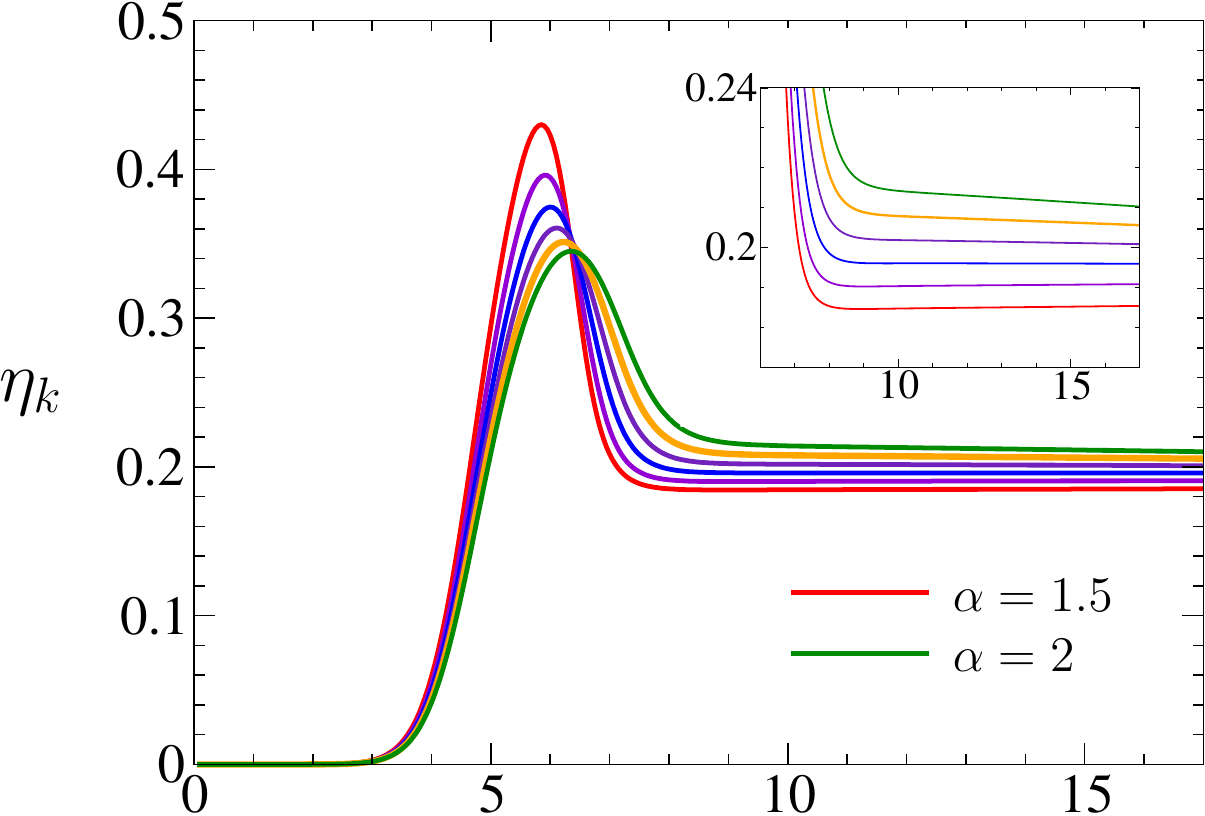}}
\vspace{0.25cm} 
\centerline{\includegraphics[width=7cm]{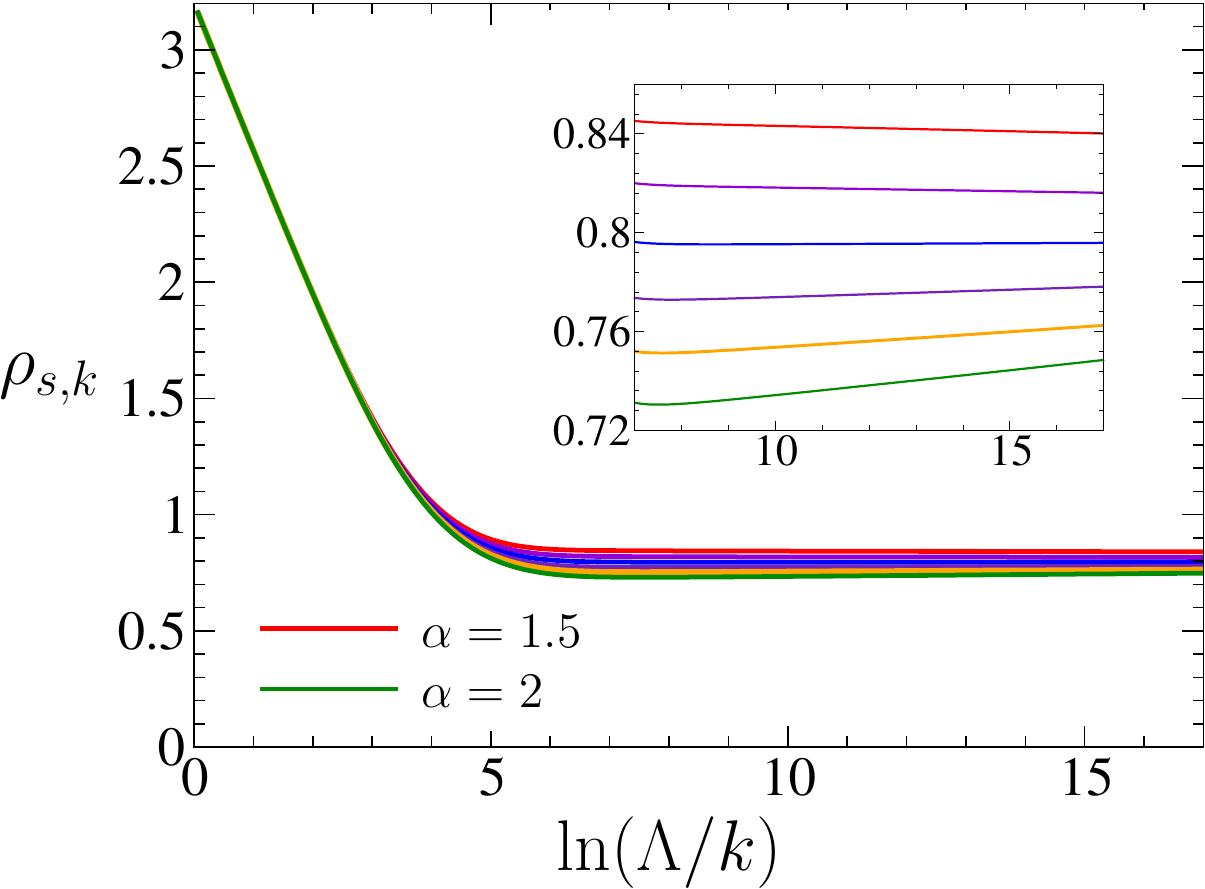}}
\caption{(Color online) $\eta_k$ and $\rho_{s,k}$ in the low-temperature phase for $\alpha=1.5,1.6,1.7,1.8,1.9,2$ (from bottom to top in the inset of the top figure and the reverse in the bottom inset) and a fixed temperature $T<\Tkt$ corresponding to $r_0=-0.0016$ (all figures are obtained with $u_0/\Lambda^2=0.003$ and $\Lambda=1$).}
\label{fig_cutoff_choice} 
\end{figure} 

In the two-dimensional case we numerically integrate the flow equations of $\tilde{U}_k', \tilde{Z}_k,\tilde{Y}_k$ starting at scale $k=\Lambda$ with the bare action (\ref{action1}) and for various values of $r_0\propto T-T_0$ (we take $u_0/\Lambda^2=0.003$). From the behavior of the RG flow in the $k\to 0$ limit, we can clearly identify a high-temperature and a low-temperature phase. The high-temperature phase is characterized by the vanishing of $\rho_{s,k}$ and $\eta_k$ at a nonzero value of $k$. In the low-temperature phase, both $\rho_{s,k}$ and $\eta_k$ remain finite for $k\to 0$ while $\rho_{0,k}$ vanishes as a power law as anticipated in the preceding section. The transition temperature $\Tkt$ is defined from the critical value $r_{0c}$ separating the two phases.

In the low-temperature phase and at fixed $\trhor$, the long-distance behavior of the RG flow depends on the infrared regulator (i.e. the parameter $\alpha$ in~(\ref{Rdef})) in a crucial way. We expect the RG trajectory to flow into a fixed point as in the standard KT theory. $\eta_k$ and $\rho_{s,k}$ (and more generally the functions $\tilde U_k(\trho)$, $\tilde Z_k(\trho)$ and $\tilde Y_k(\trho)$) should then become $k$ independent for sufficiently small $k$. 
Figure~\ref{fig_cutoff_choice} shows that for an arbitrary value of $\alpha$, in general we do not reach a fixed 
point, and $\rho_{s,k}$ and $\eta_k$ exhibit only quasi-plateaus at small $k$ with slopes that are either positive or negative
depending on $\alpha$. Thus, for each temperature $T<T_{KT}$ (but not too small, see below), it is possible to fine tune $\alpha$ such that we obtain a true plateau. We view this particular value $\aopt\equiv\aopt(T)$ as the optimal choice of the regulator. We find $\aopt(\Tkt)=2.0$ and $\aopt(T)<2$ for $T<\Tkt$. In the high-temperature phase, we take $\aopt=2$. In the following sections, we shall always consider the optimal regulators. The fact that $\aopt$ changes with $T$ is a limitation of the derivative expansion used to 
solve the flow equation~(\ref{rgeq}). In the exact solution, we expect the RG flow to reach a fixed point in the low-temperature phase regardless of the choice of the regulator. It should be noted however that a nonoptimal choice ($\alpha\neq\aopt$) leads to essentially the same long-distance physics even though there is no fixed point. In particular, the system exhibits algebraic order (except perhaps at extremely large length scales). The ultimate fate of $\rho_{s,k}$ and $\eta_k$ as $k\to 0$ (which depends on the sign of the slope of the quasi-plateau) is clearly irrelevant at macroscopic length scales of interest~\cite{note1}. 

The optimal value $\aopt\equiv\aopt(\trhor)$ depends on $\trhor$ but the universal features of the KT transition are independent of the choice of $(\trhor,\aopt(\trhor))$. In the low-temperature phase, when $\trhor$ is too large the propagator $G_k=(\Gamma_k^{(2)}+R_k)^{-1}$ does not remain positive definite due to the appearance of a pole at finite $k$, and the RG flow cannot be continued to lower $k$~\cite{note2,note5}. The lower the temperature, the smaller the renormalization point should be. 
We find that $\trhor$ must always be smaller than the minimum $\trho_{0,k}$ of the effective potential because, otherwise, a pole in the propagator appears at finite RG time $t$. Thus, it is never possible to choose $\trhor=\trho_{0,k}$. Below a certain temperature, even with $\trhor=0$, it is not possible to avoid the appearance of a pole in the propagator. The lowest temperature that can be reached corresponds to an anomalous dimension $\eta(T)\simeq 0.17$ (obtained with $\alpha=1.45$). It should be noted however that the low-temperature regime $T\ll \Tkt$, which is dominated by spinwave 
excitations, becomes trivial when one works with the Goldstone boson (i.e. the phase of the complex field $\varphi_1+i\varphi_2$) and there is no need to use the NPRG. 

\begin{figure} 
\centerline{\includegraphics[width=7cm]{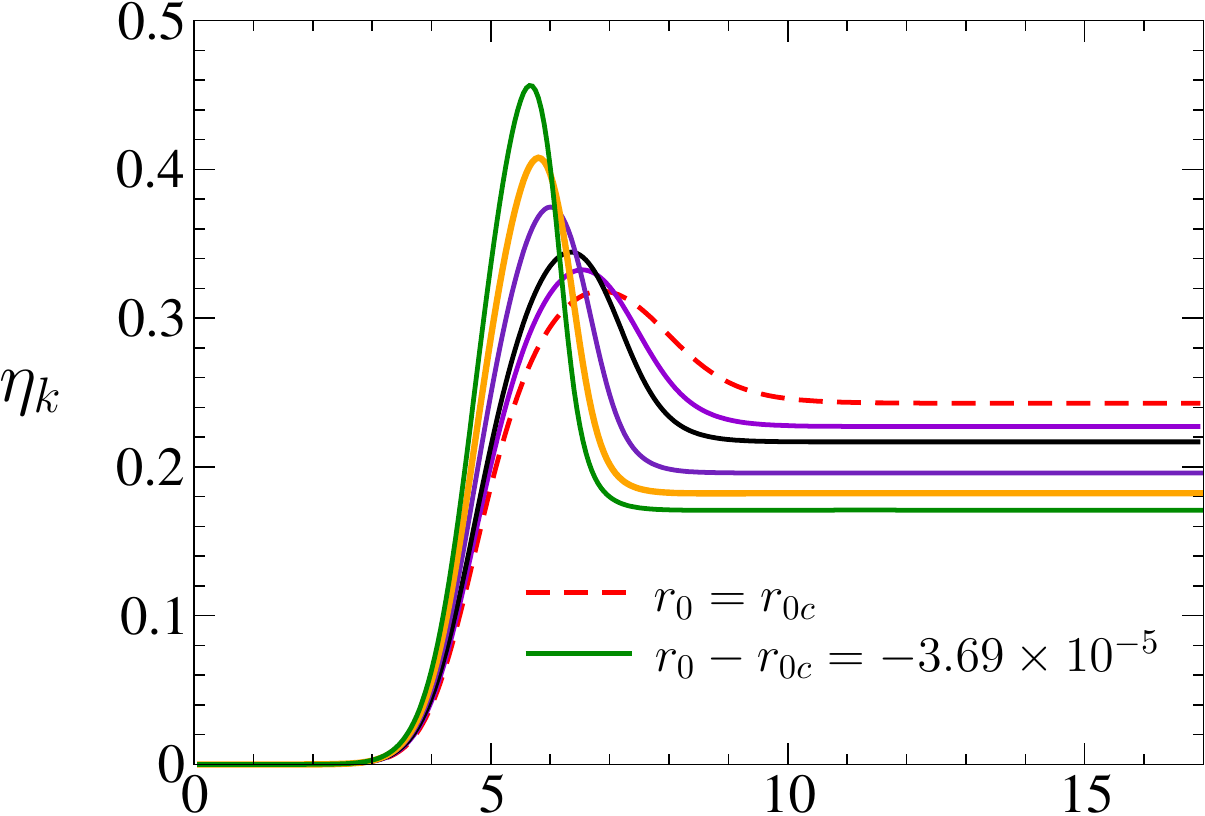}}
\vspace{0.25cm} 
\centerline{\includegraphics[width=7cm]{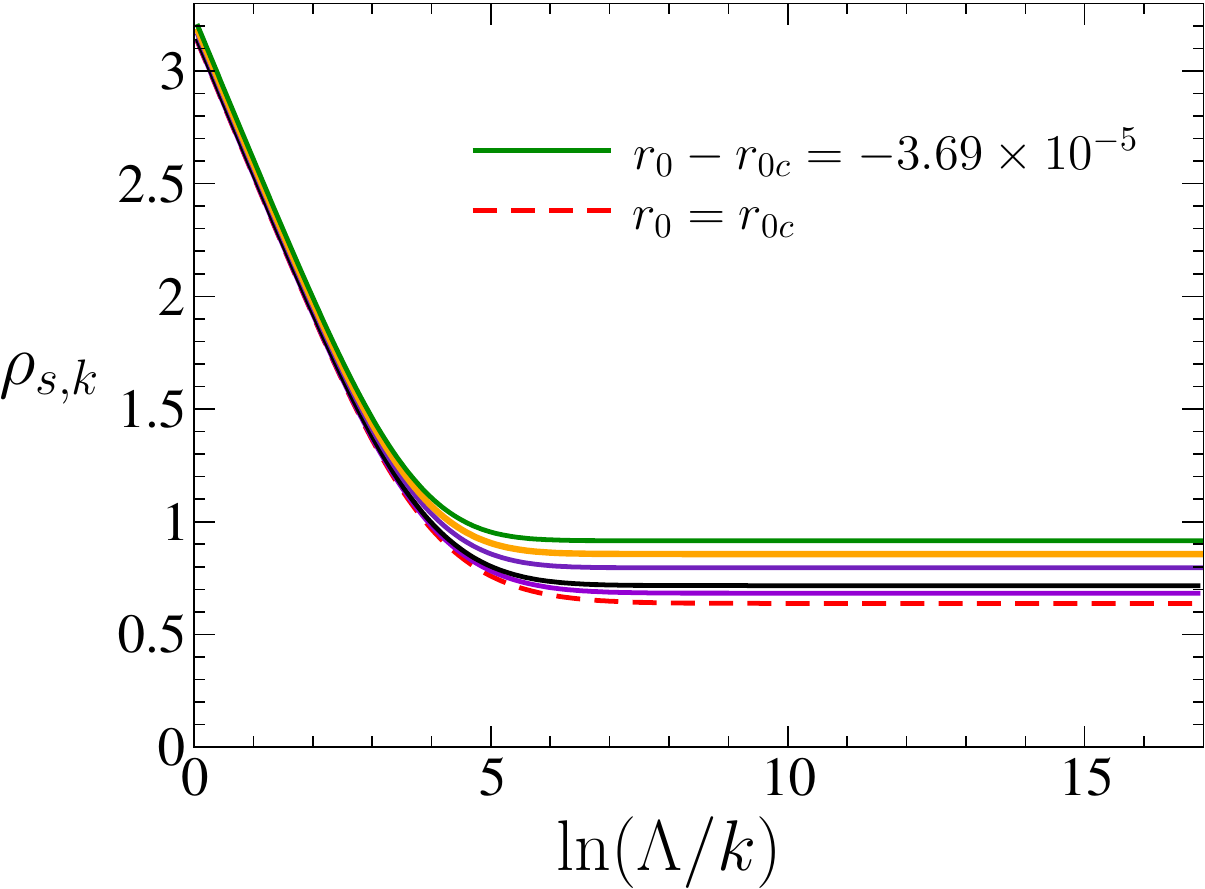}}
\caption{(Color online) Anomalous dimension $\eta_k$ and stiffness $\rho_{s,k}$ vs $\ln(\Lambda/k)$ for $\alpha=\aopt$ and various values of $r_0$ from $r_{0c}\simeq -0.0015831$ to $r_{0c}-3.69\times 10^{-5}$.}
\label{fig_rhos_eta}
\end{figure}

Figure~\ref{fig_rhos_eta} shows $\rho_{s,k}$ and $\eta_k$ for various temperatures below the KT transition temperature, obtained with the optimal parameter $\aopt$. The renormalized stiffness $\rho_s(T)=\lim_{k\to 0}\rho_{s,k}$ and the anomalous dimension $\eta(T)=\lim_{k\to 0}\eta_k$ are obtained from the plateau values of $\rho_{s,k}$ and $\eta_k$. The highest temperature for which we find a phase with a nonzero stiffness $\rho_s(T)$ provides an estimate $r_{0c}\simeq -0.0015831$ of the KT transition temperature $\Tkt$. We shall discuss other determinations of $\Tkt$ in the following sections. 

\subsection{Comparison with GW~\cite{Gersdorff01}}
\label{subsec_GW} 

GW evaluate the anomalous dimension $\eta_k$ at the flowing minimum of the effective potential while we compute it at a fixed value of the (rescaled)
field. Moreover GW do not use the prefactor $\alpha$ of the infrared regulator as a free parameter. As pointed out above, the choice $\trhor=\trho_{0,k}$ leads to the appearance of a pole in the propagator at finite $k$. We emphasize that this is not an accuracy problem
but rather an intrinsic feature of the flow equations in the derivative expansion to second order. To circumvent this difficulty, GW solve the flow equations only for a finite (scale-dependent) range of $\trho$ values around $\trho_{0,k}$~\cite{note7}. 
Even though the GW approach provides a way of computing some of the
features of the KT transition, the flow is bound to converge to the
high-temperature phase and the line of fixed points is in fact missing. In this
respect our solution is a definite improvement.

\section{KT transition} 
\label{sec_kt} 

\subsection{Suppression of amplitude fluctuations and KT physics}

\begin{figure}
\centerline{\includegraphics[width=6cm]{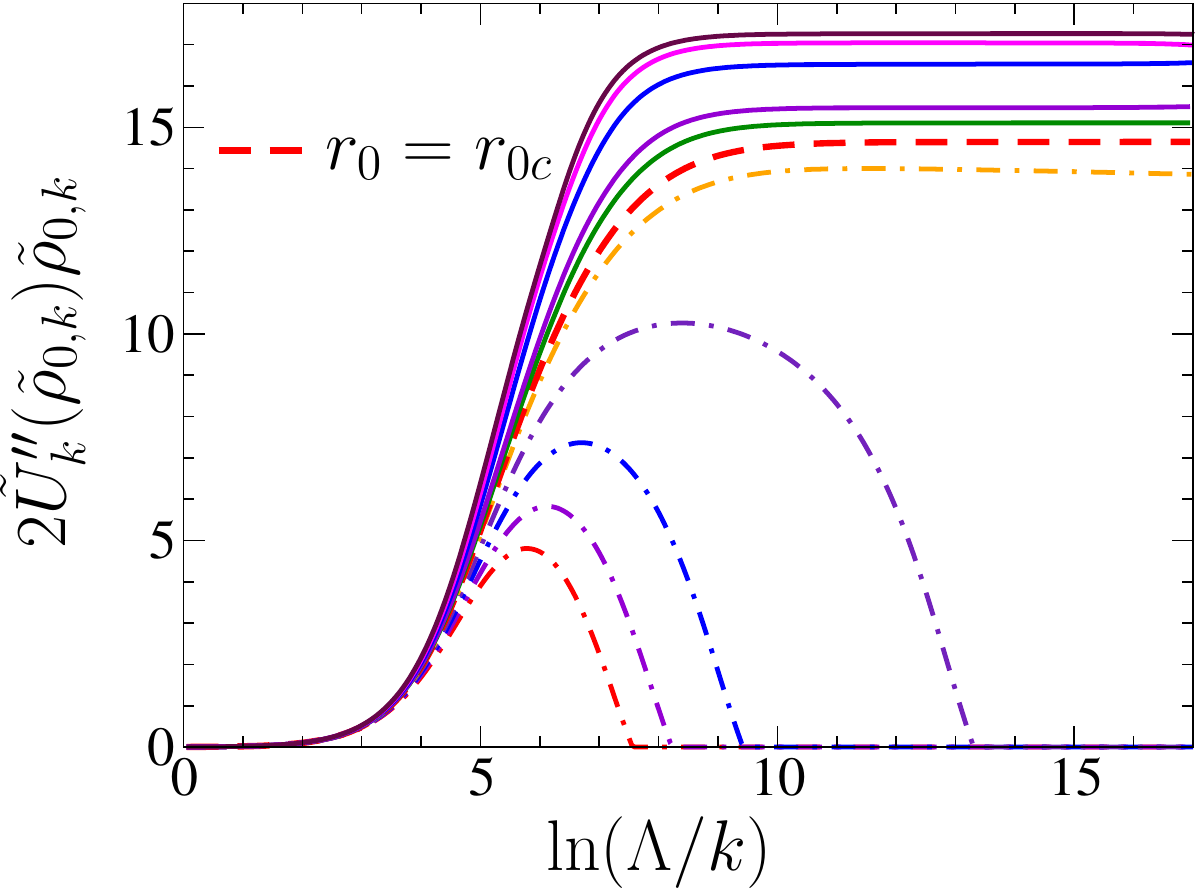}}
\caption{(Color online) Longitudinal square mass $2\tilde U''_k(\trho_{0,k})\trho_{0,k}$ vs $\ln(\Lambda/k)$ for various values of $r_0-r_{0c}$ in the low-$T$ ($-0.00162\leq r_0\leq r_{0c}$, solid lines) and high-$T$ ($r_{0c}\leq r_0\leq -0.00154$, dot-dashed lines) phases.}
\label{fig_mL} 
\end{figure}

Our results show that in the low-temperature phase as well as in the high-temperature phase in the vicinity of the KT transition, the dimensionless square ``mass'' $2\trho_{0,k}\tilde U''_k(\trho_{0,k})$ of the longitudinal mode [Eq.~(\ref{GLT})] becomes much larger than unity (Fig.~\ref{fig_mL}). For $k$ smaller than the characteristic momentum scale $k_c$ defined by $2\trho_{0,k_c}\tilde U''_{k_c}(\trho_{0,k_c}) \sim 1$, amplitude fluctuations of the two-component vector field $\varphibf$ are strongly suppressed and the flow is primarily controlled by direction fluctuations~\cite{note3}. In this long-distance regime, we expect the physics of the linear O(2) model to be similar to that of the XY model, i.e. dominated by spinwaves and vortex excitations. Note that we do not expect amplitude fluctuations to be completely frozen because this would prevent the formation of vortices since in our continuum model the field vanishes at the center of the vortex.

\subsection{Low-temperature phase}
\label{subsec_low_T} 

\begin{figure}
\centerline{\includegraphics[width=7.4cm]{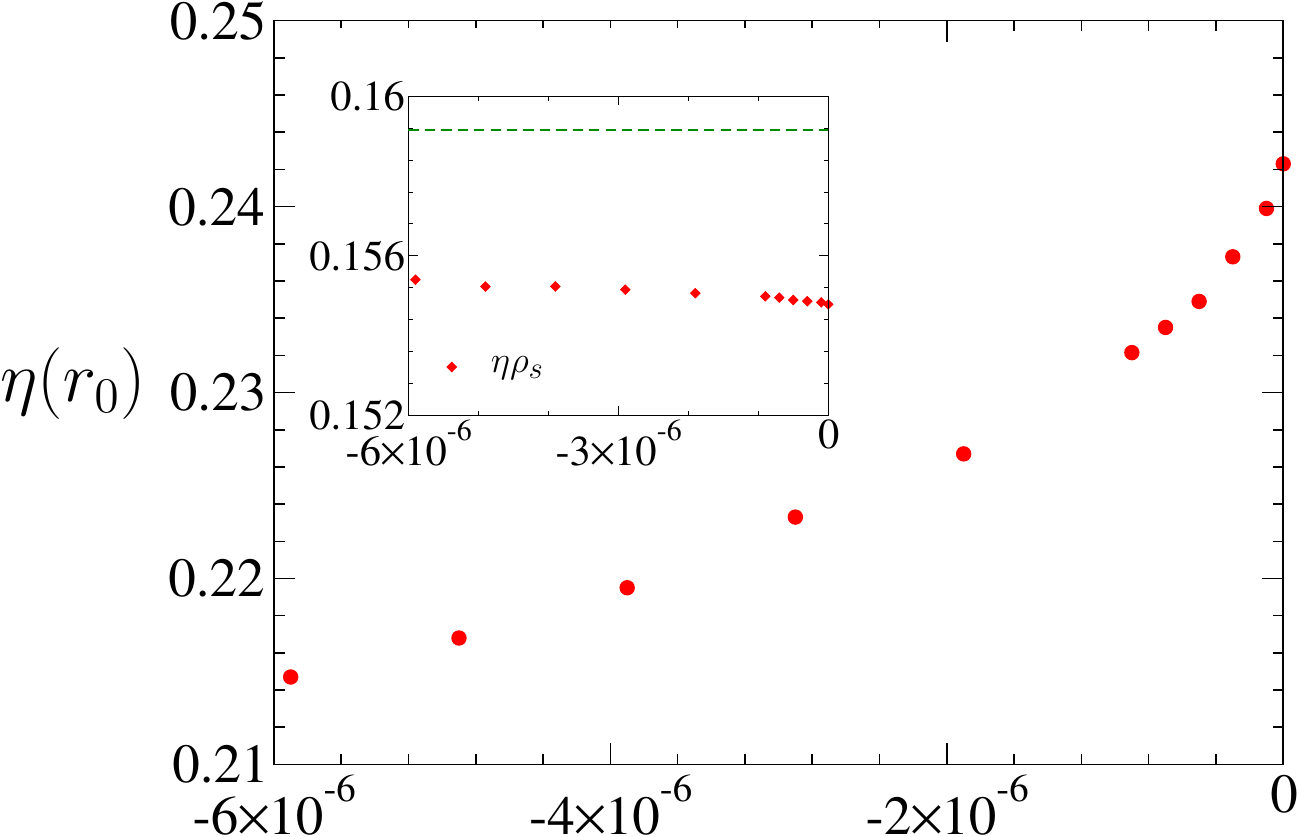}}
\vspace{0.25cm} 
\centerline{\hspace{0.35cm}\includegraphics[width=7.cm]{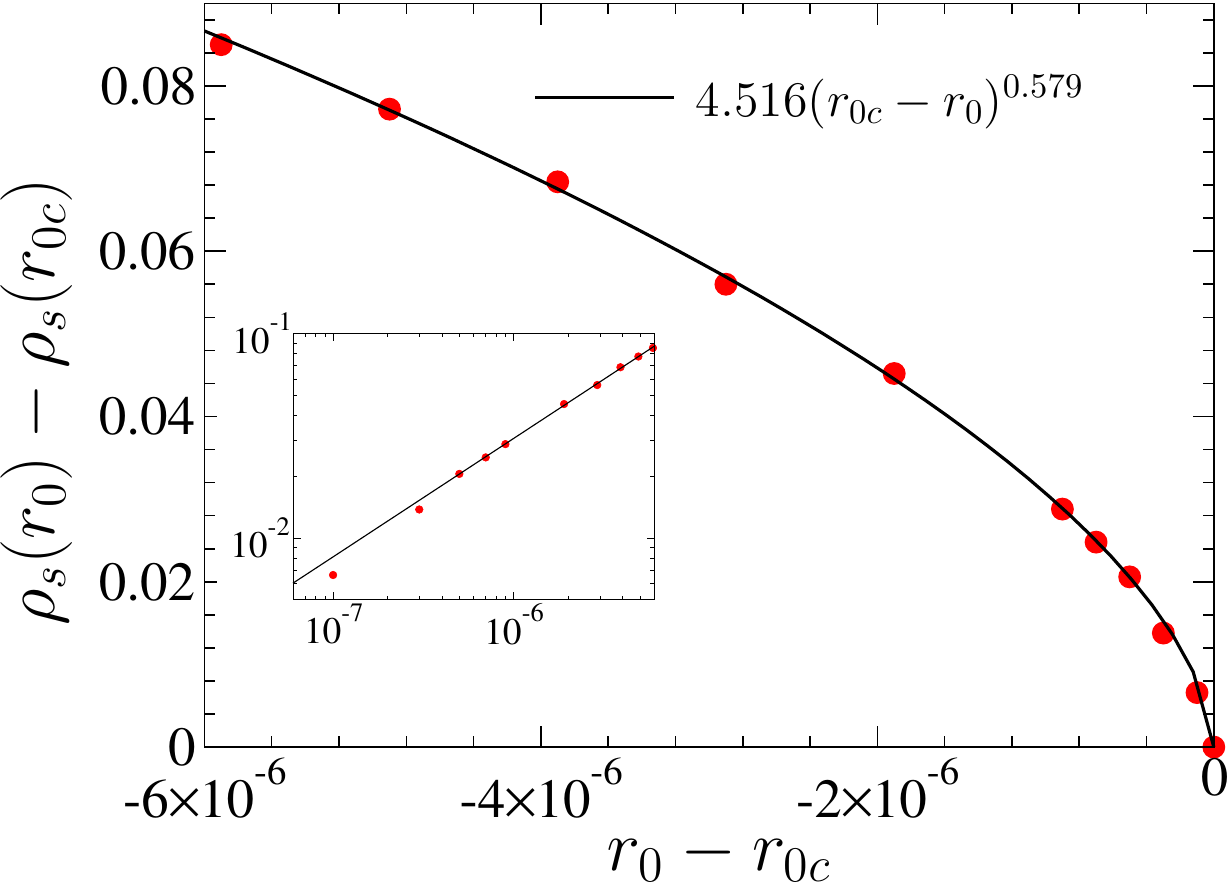}}
\caption{(Color online) Top: anomalous dimension $\eta$ vs $r_0-r_{0c}$ in the low-temperature phase for $\alpha=\aopt$. The inset shows the product $\eta\rho_s$ together with $1/2\pi$ (dashed line). 
Bottom: stiffness $\rho_s$ vs $r_0-r_{0c}$ in the low-temperature phase for $\alpha=\aopt$. The straight line in the inset shows the best power-law fit (with an exponent 0.579) in a logarithmic plot.}
\label{fig_rhos_eta_vs_T}
\end{figure}

In this section we discuss the results in the low-temperature phase $T\leq\Tkt$. 
From the numerical results obtained for $r_0\to r_{0c}\simeq -0.0015831$ (Fig.~\ref{fig_rhos_eta}), we deduce 
\begin{equation}
\rho_s(\Tkt^-) \simeq 0.64, \qquad \eta(\Tkt^-)\simeq 0.24 , 
\end{equation}
in very good agreement with the exact result $\rho_s(\Tkt^-)=2/\pi\simeq 0.6366$ and $\eta(\Tkt^-)=1/4$. By changing the initial value of $u_0$ or including a $(\varphibf^2)^3$ term in the action~(\ref{action1}), we have verified that these results are independent of the initial conditions. 

Figure~\ref{fig_rhos_eta_vs_T} shows $\rho_s(T)$ and $\eta(T)$ for $T\leq\Tkt$. The results are compatible with the temperature dependence
\begin{equation}
\rho_s(T) = \rho_s(r_{0c}^-) [ 1 + b \sqrt{r_{0c}-r_0} ] 
\label{rhos_lowT} 
\end{equation}
(see Appendix~\ref{app_bkt_flow}) of the stiffness in the vicinity of the transition even though the square-root singularity is not perfectly captured (we find the exponent 0.579 instead of 0.5). Furthermore we obtain  
\begin{equation}
\eta(T) \simeq \frac{0.155}{\rho_s(T)} .
\label{fig_etavsrhos} 
\end{equation}
In the KT theory the long-distance physics is fully determined by noninteracting spinwaves with renormalized stiffness $\rho_s(T)$ since vortices are irrelevant in the low-temperature phase. This leads to $\eta(T)=1/2\pi\rho_s(T)\simeq 0.159/\rho_s(T)$~\cite{note4}. This exact relation is well approximated by the NPRG result~(\ref{fig_etavsrhos}).

\begin{figure}
\centerline{\includegraphics[width=8.5cm]{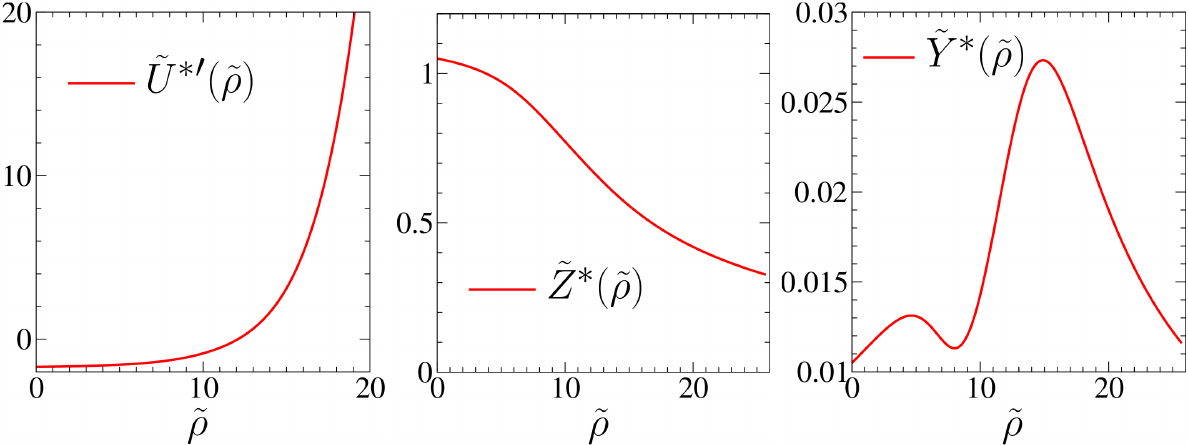}}
\caption{(Color online) Fixed-point functions $\tilde U^*{}'(\trho)$ (derivative of the effective potential), $\tilde Z^*(\trho)$ and $\tilde Y^*(\trho)$ in the low-temperature phase ($\eta=0.238$ and $\alpha=1.9$).} 
\label{fig_UZY} 
\end{figure}

\begin{figure}
\centerline{\includegraphics[width=7cm]{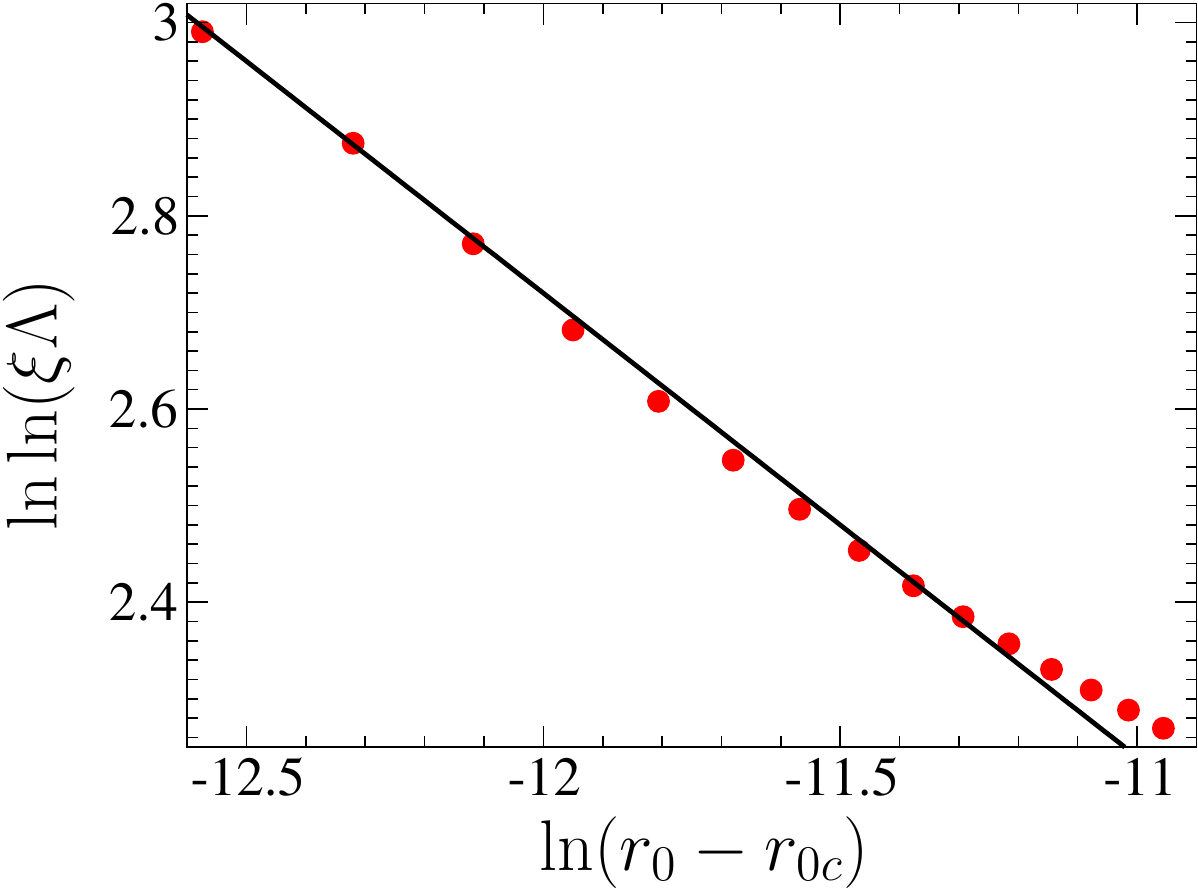}}
\caption{(Color online) $\ln\ln(\Lambda\xi)$ vs $\ln(r_0-r_{0c})$ in the high-temperature phase. The solid line corresponds to a fit of the form~(\ref{ess_scaling}) with $\gamma\simeq 0.48$.}
\label{fig_ess_scaling}
\end{figure}

Figure~\ref{fig_UZY} shows the fixed-point solutions $\tilde U^*{}'(\trho)$, $\tilde Z^*(\trho)$ and $\tilde Y^*(\trho)$  in the low-temperature phase ($\eta(T)=0.238$ and $\aopt=1.9$).

\subsection{High-temperature phase and essential scaling}
\label{subsec_high_T}

In the high-temperature phase, the propagator takes the form 
\begin{equation}
G_{k=0}(\p,\rho=0)=[Z_{k=0}(0)\p^2+U_{k=0}'(0)]^{-1}
\end{equation}
in the equilibrium field configuration $\rho=0$.
The correlation length is given by 
\begin{equation}
\xi = \left(\frac{U_{k=0}'(0)}{Z_{k=0}(0)} \right)^{-1/2} 
\end{equation}
and is shown in Fig.~\ref{fig_ess_scaling}. The best fit of the form 
\begin{equation}
\xi \sim \Lambda^{-1} \exp \left(\frac{c}{(r_0-r_{0c})^\gamma} \right)  
\label{ess_scaling}
\end{equation}
gives $\gamma\simeq 0.48$ and $c\simeq 0.048$. This result is in good agreement with the essential singularity predicted by KT theory (which corresponds to $\gamma=0.5$). Equation~(\ref{ess_scaling}) allows us to obtain $r_{0c}\simeq -0.00157746$, to be compared with the previous estimate $r_{0c}\simeq -0.0015831$ (Sec.~\ref{subsec_num2D}). KT theory predicts the product $bc=\pi/2$ to be universal (Appendix~\ref{app_bkt_flow})~\cite{note9}. The NPRG results give $bc\simeq 0.34$ (see next section for an additional comment).

\subsection{Comparison with standard KT flow}

\begin{figure}
\centerline{\includegraphics[width=5.cm,clip]{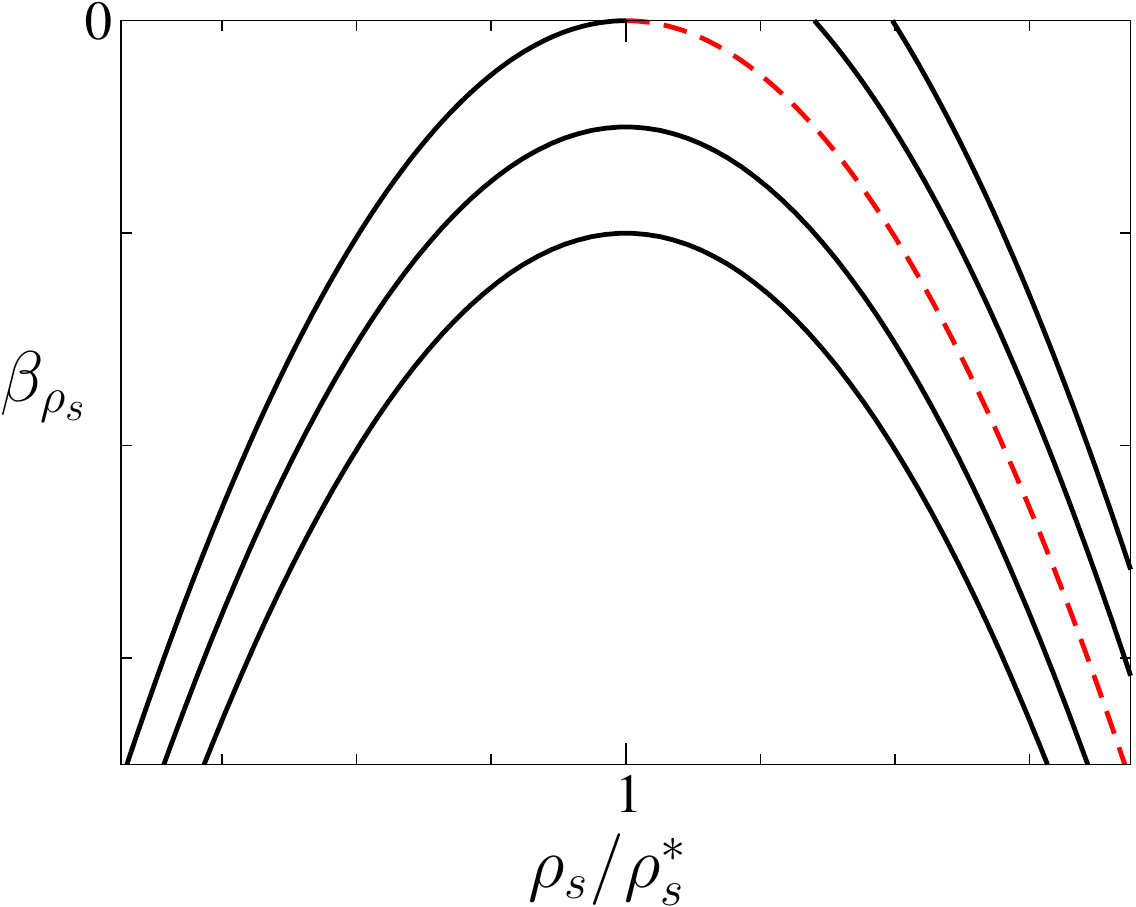}}
\vspace{-0.25cm}
\caption{(Color online) Beta function $\beta_{\rho_s}$ vs $\rho_s$ for various values of the initial fugacity $y$ obtained from the standard KT theory (Appendix~\ref{app_bkt_flow}). The dashed (red) line shows the critical line corresponding to $T=\Tkt$.}
\label{fig_flow_kt_standard_2}
\vspace{0.25cm}
\centerline{\includegraphics[width=6.5cm]{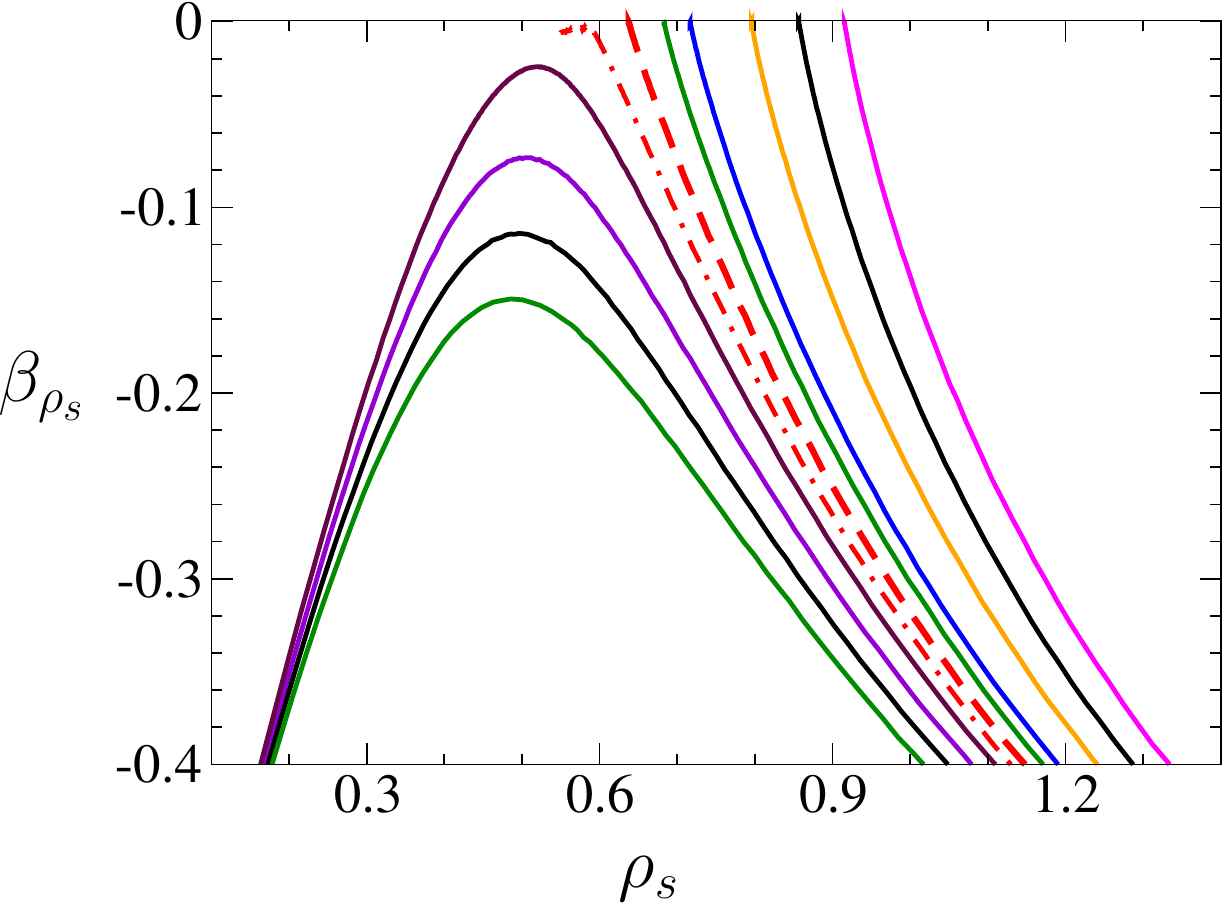}}
\vspace{-0.25cm}
\caption{(Color online) Beta function $\beta_{\rho_s}=-k d\rho_s/dk$ vs $\rho_s$ obtained from the NPRG for various initial conditions $-0.00162\leq r_0\leq -0.00154$. The dashed line 
corresponds to the critical value $r_0=-0.0015831$ deduced from the $k\to 0$ limit of $\rho_{s,k}$ (Sec.~\ref{subsec_num2D}) while the dot-dashed line corresponds to the critical value $r_0=-0.001577$ deduced from the essential scaling of the correlation length in the high-temperature phase (Sec.~\ref{subsec_high_T}).}
\label{fig_beta_rhos}
\vspace{0.25cm}
\centerline{\includegraphics[width=6.5cm]{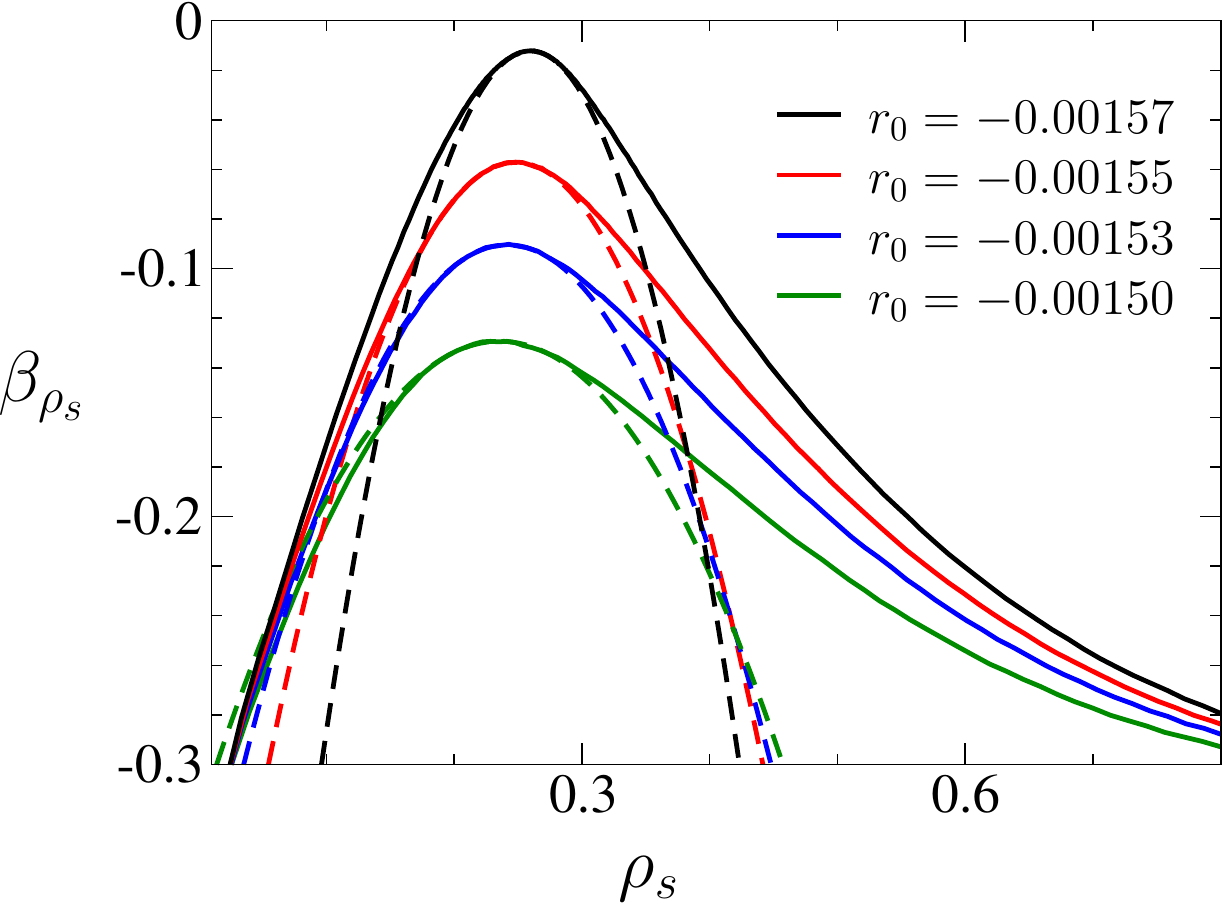}}
\vspace{-0.25cm}
\caption{(Color online) Beta function $\beta_{\rho_s}$ in the high-temperature phase ($r_0>r_{0c}$) obtained from the NPRG. The dashed lines correspond to parabolic fits.}
\label{fig_beta_rhos_fit}
\end{figure}

In the standard KT theory, the two variables of interest are the stiffness $\rho_s$ and the vortex fugacity $y$ (Appendix~\ref{app_bkt_flow}). In the NPRG approach, we follow an infinite number of variables (through the functions $U_k(\rho)$, $Z_k(\rho)$ and $Y_k(\rho)$) but we do not have access to the vortex fugacity. To make a comparison with the standard KT flow possible, we consider the beta function $\beta_{\rho_s}=-kd\rho_s/dk$. The KT theory predicts 
\begin{equation}
\beta_{\rho_s} = - \pi (\rho_s-\rho_s^*)^2 - \frac{4}{\pi}C 
\label{beta_rhos}
\end{equation}
in the vicinity of $\rho_s^*=2/\pi$, where $C$ is a constant which corresponds to different values of the initial vortex fugacity and is expected to vary linearly with $T-\Tkt$ (i.e. $r_0-r_{0c}$) (see Fig.~\ref{fig_flow_kt_standard_2} and Appendix~\ref{app_bkt_flow}). The NPRG result is shown in Figs.~\ref{fig_beta_rhos} and \ref{fig_beta_rhos_fit}. We obtain a qualitative agreement with KT theory. In particular, in the high-temperature phase $\beta_{\rho_s}$ vs $\rho_s$ is given by parabolas whose distance to the axis $\beta_{\rho_s}=0$ varies linearly with $r_0-r_{0c}$ (Fig.~\ref{fig_beta_rhos_fit}).
There are however two differences with the predictions of KT theory. First, the maxima of the parabolas are not located at $\rho_s^*$ but are shifted to lower values as $r_0$ increases. This shift indicates that $\beta_{\rho_s}$ contains a linear term in addition to the quadratic one. This linear term is however sufficiently small not to affect the essential scaling except maybe for extremely large values of the correlation length (Sec.~\ref{subsec_high_T}). Second, the prefactor of the quadratic term varies with $r_0$ and differs from $-\pi$. For the high-temperature curves shown in Fig.~\ref{fig_beta_rhos_fit}, we find that it varies between -10.76 and -3.48 as we go away from $\Tkt$. The fact that we do not find the expected prefactor $-\pi$ for the quadratic term (in particular close to the KT transition temperature) explains 
the failure to obtain the correct universal value of the product $bc$ (see preceding section). Very close to the transition point, the beta function $\beta_{\rho_s}$ looses its parabolic shape and its maximum seems to turn into a singular point. This region however corresponds to a very large renormalization time $|t|=\ln(\Lambda/k)$ and therefore an extremely large correlation length, and we do not expect our results to be fully reliable.

\section{Conclusion} 

The NPRG has proven to be a powerful method to study the critical behavior of the $d$-dimensional O($N$) model~\cite{Berges02,Delamotte12}. In this paper we have shown that this is true also in the special case $d=2$ and $N=2$: the NPRG results turn out to be in very good agreement, both qualitatively and quantitatively, with the universal features of the KT transition. 

While the KT transition has been understood for a long time in the framework of the XY model, its investigation in real materials raises new questions that are not easily answered from the standard theory: the role of the vortex core energy or the third dimension in layered quasi-twodimensional systems~\cite{Benfatto13}, the effect of disorder, the KT transition in the two-dimensional Bose gas, the coupling to fermionic degrees of freedom, etc. We believe that the NPRG can provide us with new insights into these difficult problems. In particular we would like to stress that the NPRG approach is by no means restricted to 
$\varphi^4$-type theories. This fact is of pronounced importance when one aims at analyzing effective bosonic theories derived from microscopic 
models, which naturally arise for example in the context of superconductivity~\cite{Strack08,Obert13,Strack14,Eberlein13a,Eberlein14,Friederich10,Friederich11}. In this case an expansion of the action to finite order may be
ill-defined~\cite{Strack14} and one must keep terms to all orders in the field. We emphasize that the NPRG approach also enables the computation of
non-universal properties of the system, retaining the link to the underlying microscopic model. Progress in this direction in the context of two-dimensional fermionic superfluids is underway.

\begin{acknowledgments}
We thank Walter Metzner, Marek Napi\'orkowski and Philipp Strack for useful discussions, and G. v. Gersdorff for providing us with his Diplomarbeit thesis~\cite{Gersdorff00}. PJ acknowledges support from the Polish Ministry of Science and Higher Education 
via grant IP2012 014572. Universit\'e Pierre et Marie Curie is part of Sorbonne Universit\'es. 
\end{acknowledgments}

\appendix 

\section{NPRG flow equations} 
\label{app_rgeq} 

For the sake of generality we consider the O($n$) model. 
The flow equations for the dimensionless variables read
\begin{widetext} 
\begin{align}
\dt \tilde U_k' ={}& (\eta_k-2) \tilde U_k' + \eta_k \trho \tilde U_k'' 
  -2 (n-1) L(1,0,d) \tilde{U}_k''-2 (n-1) L(1,0,d+2) \tilde{Z}_k'
  +L(0,1,d) \left(-4 \tilde{\rho } \tilde{U}_k{}^{(3)}-6
   \tilde{U}_k''\right)\nonumber \\ 
 &   -2 L(0,1,d+2) \left(\tilde{\rho } \tilde{Y}_k'+\tilde{Y}_k+\tilde{Z}_k'\right) , \label{app1} \\
\dt \tilde Z_k ={}& \eta \tilde Z_k- (-d-\eta +2)\trho \tilde Z_k' 
-2 L(1,0,d) \left((n-1) \tilde{Z}_k'+\tilde{Y}_k\right)+16
   \tilde{\rho } L(1,1,d) \tilde{U}_k'' \tilde{Z}_k'+\frac{8
   \tilde{\rho } L(1,1,d+2) \tilde{Z}_k' \left(d
   \tilde{Y}_k+\tilde{Z}_k'\right)}{d} \nonumber \\ 
 &   -2 L(0,1,d) \left(2
   \tilde{\rho } \tilde{Z}_k''+\tilde{Z}_k'\right)-\frac{16
   \tilde{\rho } \left(\tilde{U}_k''\right){}^2
   M_{\text{LT}}(2,2,d)}{d}-\frac{16 \tilde{\rho } \tilde{Y}_k
   \tilde{U}_k'' M_{\text{LT}}(2,2,d+2)}{d}-\frac{4 \tilde{\rho }
   \tilde{Y}_k^2 M_{\text{LT}}(2,2,d+4)}{d} \nonumber \\ 
 &   +\frac{16 \tilde{\rho
   } N(2,1,d) \tilde{U}_k'' \left(\tilde{Y}_k-2
   \tilde{Z}_k'\right)}{d}+\frac{8 \tilde{\rho } \tilde{Y}_k
   N(2,1,d+2) \left(\tilde{Y}_k-2 \tilde{Z}_k'\right)}{d} \label{app2},
\end{align} 
and 
\begin{align} 
\dt \tilde Y_k ={}& (d-2+2\eta) \tilde Y_k- (-d-\eta +2)\trho \tilde Y_k'  -\frac{8 \tilde{Y}_k N(2,1,d+2)
   \left(\tilde{Y}_k-2 \tilde{Z}_k'\right)}{d}-\frac{16
   N(2,1,d) \tilde{U}_k'' \left(\tilde{Y}_k-2
   \tilde{Z}_k'\right)}{d} \nonumber \\ 
 &+4 (n-1) \tilde{Y}_k L(2,0,d) \tilde{U}_k''+2 L(1,0,d)
   \left(\frac{\tilde{Y}_k}{\tilde{\rho }}-(n-1)
   \tilde{Y}_k'\right)+\frac{4 (n-1) L(2,0,d+2) \tilde{Z}_k'
   \left(d \tilde{Y}_k+\tilde{Z}_k'\right)}{d} \nonumber \\ 
 &  +8 L(0,2,d) \left(2
   \tilde{\rho } \tilde{U}_k{}^{(3)}+3 \tilde{U}_k''\right)
   \left(\tilde{\rho }
   \tilde{Y}_k'+\tilde{Y}_k+\tilde{Z}_k'\right)-16 L(1,1,d)
   \tilde{U}_k'' \tilde{Z}_k'-\frac{2 L(0,1,d) \left(5
   \tilde{\rho } \tilde{Y}_k'+2 \tilde{\rho }^2
   \tilde{Y}_k''+\tilde{Y}_k\right)}{\tilde{\rho }} \nonumber \\ 
 &  +\frac{4 (2
   d+1) L(0,2,d+2) \left(\tilde{\rho }
   \tilde{Y}_k'+\tilde{Y}_k+\tilde{Z}_k'\right){}^2}{d}-\frac{8
   L(1,1,d+2) \tilde{Z}_k' \left(d
   \tilde{Y}_k+\tilde{Z}_k'\right)}{d}-\frac{8 M_{\rm L}(0,4,d)
   \left(2 \tilde{\rho } \tilde{U}_k{}^{(3)}+3
   \tilde{U}_k''\right){}^2}{d} \nonumber \\ 
 &  -\frac{16 M_{\rm L}(0,4,d+2)
   \left(2 \tilde{\rho } \tilde{U}_k{}^{(3)}+3
   \tilde{U}_k''\right) \left(\tilde{\rho }
   \tilde{Y}_k'+\tilde{Y}_k+\tilde{Z}_k'\right)}{d}-\frac{8
   M_{\rm L}(0,4,d+4) \left(\tilde{\rho }
   \tilde{Y}_k'+\tilde{Y}_k+\tilde{Z}_k'\right){}^2}{d} \nonumber \\ 
 &   +\frac{16
   \tilde{Y}_k M_{\rm LT}(2,2,d+2) \tilde{U}_k''}{d}+\frac{16
   M_{\rm LT}(2,2,d) \left(\tilde{U}_k''\right){}^2}{d}+\frac{4
   \tilde{Y}_k^2 M_{\rm LT}(2,2,d+4)}{d}-\frac{16 (n-1)
   M_{\rm T}(4,0,d+2) \tilde{U}_k'' \tilde{Z}_k'}{d} \nonumber \\ 
 &   -\frac{8
   (n-1) M_{\rm T}(4,0,d)
   \left(\tilde{U}_k''\right){}^2}{d}-\frac{8 (n-1)
   M_{\rm T}(4,0,d+4) \left(\tilde{Z}_k'\right){}^2}{d} .\label{app3}
\end{align}
\end{widetext} 
Note that with the choice of normalization in Eqs.~(\ref{dimless}), the angular factor $v_d$ does not appear in the flow equations. Equations~(\ref{app1}-\ref{app3}) must be solved with the initial conditions $\tilde U_\Lambda(\trho)=\tilde r_0\trho + (\tilde u_0/6)\trho^2$, $\tilde Z_\Lambda(\trho)=1$ and $\tilde Y_\Lambda(\trho)=0$, where $\tilde r_0=r_0\Lambda^{-2}$ and $\tilde u_0=u_0v_d\Lambda^{d-4}$.

The equation for $\eta_k$ is obtained from the renormalization condition $\dt \tilde Z_k(\trhor)=0$. 
We have introduced the threshold functions 
\begin{equation}
\begin{split}
L(n_1,n_2,d) &= - \half \tdt \int_0^\infty dy\, y^{d/2-1} \tilde G^{n_1}_{\rm T} \tilde G^{n_2}_{\rm L} , \\
M_{\rm T}(n_1,n_2,d) &= - \half \tdt \int_0^\infty dy\, y^{d/2} \tilde G'_{\rm T}\tilde G^{n_1-4}_{\rm T} \tilde G^{n_2}_{\rm L} , \\
M_{\rm L}(n_1,n_2,d) &= - \half \tdt \int_0^\infty dy\, y^{d/2} \tilde G'_{\rm L}\tilde G^{n_1}_{\rm T} \tilde G^{n_2-4}_{\rm L} , \\
M_{\rm LT}(n_1,n_2,d) &= - \half \tdt \int_0^\infty dy\, y^{d/2} \tilde G'_{\rm L} \tilde G'_{\rm T} \tilde G^{n_1-2}_{\rm T} \tilde G^{n_2-2}_{\rm L} , \\
N(n_1,n_2,d) &= \half \tdt \int_0^\infty dy\, y^{d/2} \tilde G'_{\rm T} \tilde G^{n_1-2}_{\rm T} \tilde G^{n_2}_{\rm L} , 
\end{split}
\end{equation}
and the dimensionless transverse and longitudinal propagators [see Eqs.~(\ref{GLT})]~\cite{note2}
\begin{equation}
\begin{split}
\tilde G_{\rm L}^{-1}(y) &= y[\tilde Z_k + \trho \tilde Y_k + r(y)] + \tilde U_k' + 2\trho \tilde U_k'' , \\    
\tilde G_{\rm T}^{-1}(y) &=  y[\tilde Z_k + r(y)] + \tilde U_k' . 
\end{split}
\end{equation}
To alleviate the notations, we do not write explicitly the $k$ and $\trho$ dependence of the threshold functions and the propagators, and use $\tilde G'=\partial_y \tilde G$. The operator $\tdt$ is defined by
\begin{equation}
\begin{split}
\tdt &= (\dt R_k) \frac{\partial}{\partial R_k}\biggl|_{R_k'} + (\dt R'_k) \frac{\partial}{\partial R'_k}\biggl|_{R_k} , \\ 
\dt R_k &= - Z_k k^2 y (\eta r+2yr') , \\ 
\dt R'_k &= - Z_k [ \eta r+(\eta+4)yr'+2y^2r'' ] ,
\end{split}
\end{equation}
with $R_k'(\q^2)=\partial_{\q^2} R_k(\q^2)=Z_k(r+yr')$, $y=\q^2/k^2$ and $r\equiv r(y)$, $r'\equiv r'(y)$, $r''\equiv r''(y)$. 

The threshold functions are used here to present the flow equations in a concise way. In practice however, we write the rhs of the flow equations as a single integral which is computed using an equally-spaced grid in the variable $\sqrt{y}$.

\section{KT RG flow} 
\label{app_bkt_flow}

\begin{figure}
\centerline{\includegraphics[width=5.cm,clip]{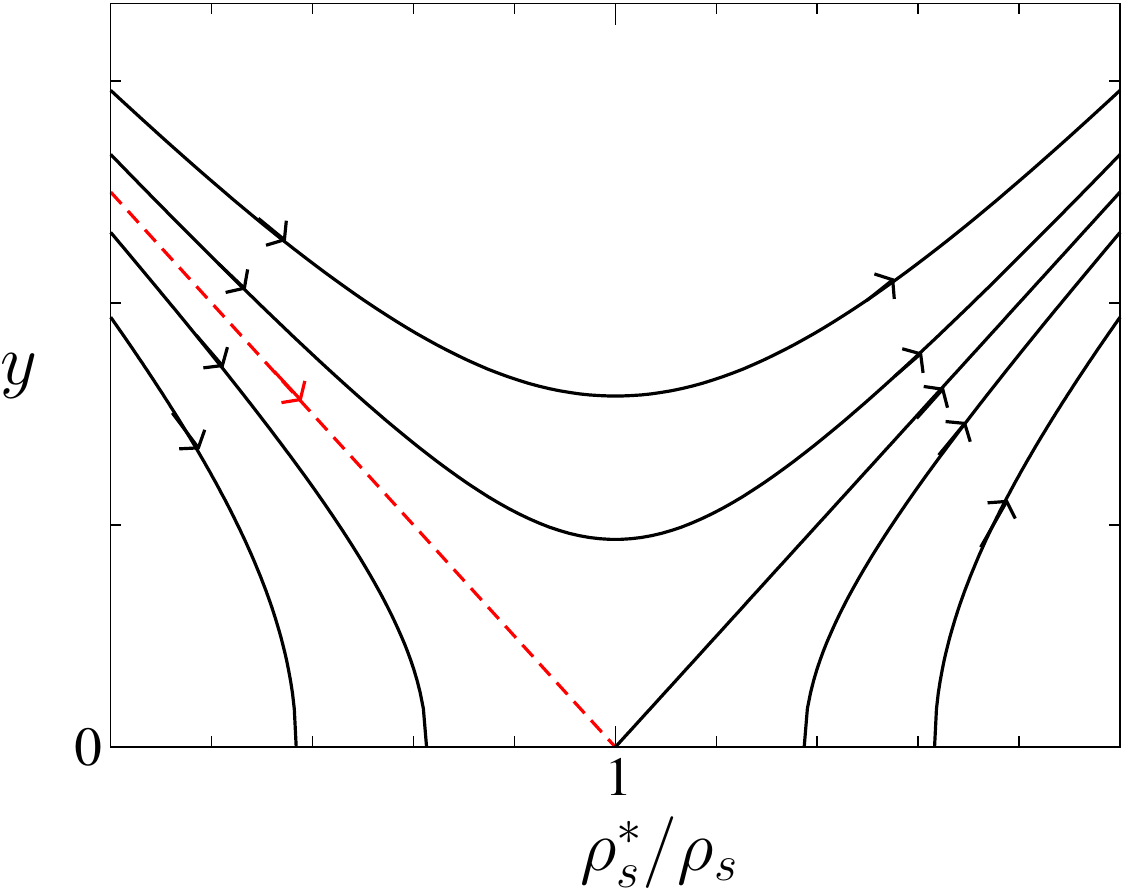}}
\caption{(Color online) Schematic KT flow in the vicinity of $y=0,\rho_s^*=2/\pi$ as obtained from Eqs.~(\ref{app_bkt1}). The dashed (red) line shows the critical line corresponding to $T=\Tkt$.}
\label{fig_flow_kt_standard_1}
\end{figure}

In the standard KT theory, the two variables of interest are the stiffness $\rho_s$ and the vortex fugacity $y$. In the limit of small fugacity, the RG equations read
\begin{equation}
\begin{split}
\frac{d\rho_s(l)^{-1}}{dl} &= 4\pi^3 y(l)^2 + \calO\bigl(y(l)^4\bigr) , \\
\frac{dy(l)}{dl} &= [2-\pi\rho_s(l)] y(l) + \calO\bigl(y(l)^3\bigr) , 
\end{split}
\label{app_bkt1}
\end{equation}
where we use the positive RG time $l=\ln(\Lamb/k)\equiv -t$ following usual notations ($\Lambda^{-1}$ denotes a microscopic length scale, e.g. the lattice spacing if we consider the XY model). The corresponding flow diagram is shown in Fig.~\ref{fig_flow_kt_standard_1}. The RG trajectories in the plane $(y,\rho_s^{-1})$ are given by the hyperbolas 
\begin{equation}
y = \frac{1}{\pi^4} \left(\rho_s^{-1}-\frac{\pi}{2} \right)^2 + \frac{C}{4\pi^2} . 
\end{equation}
The critical trajectory is defined by $C=0$ while $C>0$ and $C<0$ correspond to the high- and low-temperature phases, respectively. Assuming that $C=b^2(T-\Tkt)$ vanishes linearly with $T-\Tkt$ and integrating the flow equations~(\ref{app_bkt1}) we obtain~\cite{Chaikin_book} 
\begin{equation}
\xi(T) \sim \Lambda^{-1} \exp\left(\frac{\pi}{2b\sqrt{T-\Tkt}}\right) \qquad (T\to \Tkt^+) 
\end{equation}
in the high-temperature phase, and 
\begin{equation}
\rho_s(T) = \frac{2}{\pi} ( 1 + b \sqrt{\Tkt-T} )  \qquad (T\to \Tkt^-) 
\end{equation} 
in the low temperature phase.  

To make a comparison possible with the NPRG approach to the linear O(2) model, where the vortex fugacity $y$ is not accessible, we consider the beta function $\beta_{\rho_s} = d\rho_s(l)/dl$ versus $\rho_s(l)$. From Eqs.~(\ref{app_bkt1}), we deduce 
\begin{equation}
\beta_{\rho_s} = - \pi [\rho_s(l)-\rho_s^*]^2 - \frac{4}{\pi} C  
\end{equation}
in the vicinity of the critical point $\rho_s^*=2/\pi$ (Fig.~\ref{fig_flow_kt_standard_2}). Different curves are labeled by the constant $C$ and correspond to different values of the initial vortex fugacity $y$.



\begin{thebibliography}{48}%
\makeatletter
\providecommand \@ifxundefined [1]{%
 \@ifx{#1\undefined}
}%
\providecommand \@ifnum [1]{%
 \ifnum #1\expandafter \@firstoftwo
 \else \expandafter \@secondoftwo
 \fi
}%
\providecommand \@ifx [1]{%
 \ifx #1\expandafter \@firstoftwo
 \else \expandafter \@secondoftwo
 \fi
}%
\providecommand \natexlab [1]{#1}%
\providecommand \enquote  [1]{``#1''}%
\providecommand \bibnamefont  [1]{#1}%
\providecommand \bibfnamefont [1]{#1}%
\providecommand \citenamefont [1]{#1}%
\providecommand \href@noop [0]{\@secondoftwo}%
\providecommand \href [0]{\begingroup \@sanitize@url \@href}%
\providecommand \@href[1]{\@@startlink{#1}\@@href}%
\providecommand \@@href[1]{\endgroup#1\@@endlink}%
\providecommand \@sanitize@url [0]{\catcode `\\12\catcode `\$12\catcode
  `\&12\catcode `\#12\catcode `\^12\catcode `\_12\catcode `\%12\relax}%
\providecommand \@@startlink[1]{}%
\providecommand \@@endlink[0]{}%
\providecommand \url  [0]{\begingroup\@sanitize@url \@url }%
\providecommand \@url [1]{\endgroup\@href {#1}{\urlprefix }}%
\providecommand \urlprefix  [0]{URL }%
\providecommand \Eprint [0]{\href }%
\providecommand \doibase [0]{http://dx.doi.org/}%
\providecommand \selectlanguage [0]{\@gobble}%
\providecommand \bibinfo  [0]{\@secondoftwo}%
\providecommand \bibfield  [0]{\@secondoftwo}%
\providecommand \translation [1]{[#1]}%
\providecommand \BibitemOpen [0]{}%
\providecommand \bibitemStop [0]{}%
\providecommand \bibitemNoStop [0]{.\EOS\space}%
\providecommand \EOS [0]{\spacefactor3000\relax}%
\providecommand \BibitemShut  [1]{\csname bibitem#1\endcsname}%
\let\auto@bib@innerbib\@empty
\bibitem [{\citenamefont {Berezinskii}(1970)}]{Berezinskii70}%
  \BibitemOpen
  \bibfield  {author} {\bibinfo {author} {\bibfnamefont {V.~L.}\ \bibnamefont
  {Berezinskii}},\ }\href@noop {} {\bibfield  {journal} {\bibinfo  {journal}
  {Sov. Phys. JETP}\ }\textbf {\bibinfo {volume} {32}},\ \bibinfo {pages} {493}
  (\bibinfo {year} {1970})}\BibitemShut {NoStop}%
\bibitem [{\citenamefont {Berezinskii}(1971)}]{Berezinskii71}%
  \BibitemOpen
  \bibfield  {author} {\bibinfo {author} {\bibfnamefont {V.~L.}\ \bibnamefont
  {Berezinskii}},\ }\href@noop {} {\bibfield  {journal} {\bibinfo  {journal}
  {Sov. Phys. JETP}\ }\textbf {\bibinfo {volume} {34}},\ \bibinfo {pages} {610}
  (\bibinfo {year} {1971})}\BibitemShut {NoStop}%
\bibitem [{\citenamefont {Kosterlitz}\ and\ \citenamefont
  {Thouless}(1973)}]{Kosterlitz73}%
  \BibitemOpen
  \bibfield  {author} {\bibinfo {author} {\bibfnamefont {J.~M.}\ \bibnamefont
  {Kosterlitz}}\ and\ \bibinfo {author} {\bibfnamefont {D.~J.}\ \bibnamefont
  {Thouless}},\ }\href {\doibase doi:10.1088/0022-3719/6/7/010} {\bibfield
  {journal} {\bibinfo  {journal} {J. of Phys. C}\ }\textbf {\bibinfo {volume}
  {6}},\ \bibinfo {pages} {1181} (\bibinfo {year} {1973})}\BibitemShut
  {NoStop}%
\bibitem [{\citenamefont {Kosterlitz}\ and\ \citenamefont
  {Thouless}(1974)}]{Kosterlitz74}%
  \BibitemOpen
  \bibfield  {author} {\bibinfo {author} {\bibfnamefont {J.~M.}\ \bibnamefont
  {Kosterlitz}}\ and\ \bibinfo {author} {\bibfnamefont {D.~J.}\ \bibnamefont
  {Thouless}},\ }\href {\doibase doi:10.1088/0022-3719/7/6/005} {\bibfield
  {journal} {\bibinfo  {journal} {J. Phys. C}\ }\textbf {\bibinfo {volume}
  {7}},\ \bibinfo {pages} {1046} (\bibinfo {year} {1974})}\BibitemShut
  {NoStop}%
\bibitem [{\citenamefont {Rudnick}(1978)}]{Rudnick78}%
  \BibitemOpen
  \bibfield  {author} {\bibinfo {author} {\bibfnamefont {I.}~\bibnamefont
  {Rudnick}},\ }\href {\doibase 10.1103/PhysRevLett.40.1454} {\bibfield
  {journal} {\bibinfo  {journal} {Phys. Rev. Lett.}\ }\textbf {\bibinfo
  {volume} {40}},\ \bibinfo {pages} {1454} (\bibinfo {year}
  {1978})}\BibitemShut {NoStop}%
\bibitem [{\citenamefont {Bishop}\ and\ \citenamefont
  {Reppy}(1978)}]{Bishop78}%
  \BibitemOpen
  \bibfield  {author} {\bibinfo {author} {\bibfnamefont {D.~J.}\ \bibnamefont
  {Bishop}}\ and\ \bibinfo {author} {\bibfnamefont {J.~D.}\ \bibnamefont
  {Reppy}},\ }\href {\doibase 10.1103/PhysRevLett.40.1727} {\bibfield
  {journal} {\bibinfo  {journal} {Phys. Rev. Lett.}\ }\textbf {\bibinfo
  {volume} {40}},\ \bibinfo {pages} {1727} (\bibinfo {year}
  {1978})}\BibitemShut {NoStop}%
\bibitem [{\citenamefont {Maps}\ and\ \citenamefont {Hallock}(1981)}]{Maps81}%
  \BibitemOpen
  \bibfield  {author} {\bibinfo {author} {\bibfnamefont {J.}~\bibnamefont
  {Maps}}\ and\ \bibinfo {author} {\bibfnamefont {R.~B.}\ \bibnamefont
  {Hallock}},\ }\href {\doibase 10.1103/PhysRevLett.47.1533} {\bibfield
  {journal} {\bibinfo  {journal} {Phys. Rev. Lett.}\ }\textbf {\bibinfo
  {volume} {47}},\ \bibinfo {pages} {1533} (\bibinfo {year}
  {1981})}\BibitemShut {NoStop}%
\bibitem [{\citenamefont {Maps}\ and\ \citenamefont {Hallock}(1982)}]{Maps82}%
  \BibitemOpen
  \bibfield  {author} {\bibinfo {author} {\bibfnamefont {J.}~\bibnamefont
  {Maps}}\ and\ \bibinfo {author} {\bibfnamefont {R.~B.}\ \bibnamefont
  {Hallock}},\ }\href {\doibase 10.1103/PhysRevB.26.3979} {\bibfield  {journal}
  {\bibinfo  {journal} {Phys. Rev. B}\ }\textbf {\bibinfo {volume} {26}},\
  \bibinfo {pages} {3979} (\bibinfo {year} {1982})}\BibitemShut {NoStop}%
\bibitem [{\citenamefont {Resnick}\ \emph {et~al.}(1981)\citenamefont
  {Resnick}, \citenamefont {Garland}, \citenamefont {Boyd}, \citenamefont
  {Shoemaker},\ and\ \citenamefont {Newrock}}]{Resnick81}%
  \BibitemOpen
  \bibfield  {author} {\bibinfo {author} {\bibfnamefont {D.~J.}\ \bibnamefont
  {Resnick}}, \bibinfo {author} {\bibfnamefont {J.~C.}\ \bibnamefont
  {Garland}}, \bibinfo {author} {\bibfnamefont {J.~T.}\ \bibnamefont {Boyd}},
  \bibinfo {author} {\bibfnamefont {S.}~\bibnamefont {Shoemaker}}, \ and\
  \bibinfo {author} {\bibfnamefont {R.~S.}\ \bibnamefont {Newrock}},\ }\href
  {\doibase 10.1103/PhysRevLett.47.1542} {\bibfield  {journal} {\bibinfo
  {journal} {Phys. Rev. Lett.}\ }\textbf {\bibinfo {volume} {47}},\ \bibinfo
  {pages} {1542} (\bibinfo {year} {1981})}\BibitemShut {NoStop}%
\bibitem [{\citenamefont {Hadzibabic}\ \emph {et~al.}(2006)\citenamefont
  {Hadzibabic}, \citenamefont {Kr\"uger}, \citenamefont {Cheneau},
  \citenamefont {Battelier},\ and\ \citenamefont {Dalibard}}]{Hadzibabic06}%
  \BibitemOpen
  \bibfield  {author} {\bibinfo {author} {\bibfnamefont {Z.}~\bibnamefont
  {Hadzibabic}}, \bibinfo {author} {\bibfnamefont {P.}~\bibnamefont
  {Kr\"uger}}, \bibinfo {author} {\bibfnamefont {M.}~\bibnamefont {Cheneau}},
  \bibinfo {author} {\bibfnamefont {B.}~\bibnamefont {Battelier}}, \ and\
  \bibinfo {author} {\bibfnamefont {J.}~\bibnamefont {Dalibard}},\ }\href
  {\doibase doi:10.1038/nature04851} {\bibfield  {journal} {\bibinfo  {journal}
  {Nature}\ }\textbf {\bibinfo {volume} {441}},\ \bibinfo {pages} {1118}
  (\bibinfo {year} {2006})}\BibitemShut {NoStop}%
\bibitem [{\citenamefont {Clad\'e}\ \emph {et~al.}(2009)\citenamefont
  {Clad\'e}, \citenamefont {Ryu}, \citenamefont {Ramanathan}, \citenamefont
  {Helmerson},\ and\ \citenamefont {Phillips}}]{Clade09}%
  \BibitemOpen
  \bibfield  {author} {\bibinfo {author} {\bibfnamefont {P.}~\bibnamefont
  {Clad\'e}}, \bibinfo {author} {\bibfnamefont {C.}~\bibnamefont {Ryu}},
  \bibinfo {author} {\bibfnamefont {A.}~\bibnamefont {Ramanathan}}, \bibinfo
  {author} {\bibfnamefont {K.}~\bibnamefont {Helmerson}}, \ and\ \bibinfo
  {author} {\bibfnamefont {W.~D.}\ \bibnamefont {Phillips}},\ }\href {\doibase
  10.1103/PhysRevLett.102.170401} {\bibfield  {journal} {\bibinfo  {journal}
  {Phys. Rev. Lett.}\ }\textbf {\bibinfo {volume} {102}},\ \bibinfo {pages}
  {170401} (\bibinfo {year} {2009})}\BibitemShut {NoStop}%
\bibitem [{\citenamefont {Tung}\ \emph {et~al.}(2010)\citenamefont {Tung},
  \citenamefont {Lamporesi}, \citenamefont {Lobser}, \citenamefont {Xia},\ and\
  \citenamefont {Cornell}}]{Tung10}%
  \BibitemOpen
  \bibfield  {author} {\bibinfo {author} {\bibfnamefont {S.}~\bibnamefont
  {Tung}}, \bibinfo {author} {\bibfnamefont {G.}~\bibnamefont {Lamporesi}},
  \bibinfo {author} {\bibfnamefont {D.}~\bibnamefont {Lobser}}, \bibinfo
  {author} {\bibfnamefont {L.}~\bibnamefont {Xia}}, \ and\ \bibinfo {author}
  {\bibfnamefont {E.~A.}\ \bibnamefont {Cornell}},\ }\href {\doibase
  10.1103/PhysRevLett.105.230408} {\bibfield  {journal} {\bibinfo  {journal}
  {Phys. Rev. Lett.}\ }\textbf {\bibinfo {volume} {105}},\ \bibinfo {pages}
  {230408} (\bibinfo {year} {2010})}\BibitemShut {NoStop}%
\bibitem [{\citenamefont {Desbuquois}\ \emph {et~al.}(2012)\citenamefont
  {Desbuquois}, \citenamefont {Chomaz}, \citenamefont {Yefsah}, \citenamefont
  {L\'eonard}, \citenamefont {Beugnon}, \citenamefont {Weitenberg},\ and\
  \citenamefont {Dalibard}}]{Desbuquois12}%
  \BibitemOpen
  \bibfield  {author} {\bibinfo {author} {\bibfnamefont {R.}~\bibnamefont
  {Desbuquois}}, \bibinfo {author} {\bibfnamefont {L.}~\bibnamefont {Chomaz}},
  \bibinfo {author} {\bibfnamefont {T.}~\bibnamefont {Yefsah}}, \bibinfo
  {author} {\bibfnamefont {J.}~\bibnamefont {L\'eonard}}, \bibinfo {author}
  {\bibfnamefont {J.}~\bibnamefont {Beugnon}}, \bibinfo {author} {\bibfnamefont
  {C.}~\bibnamefont {Weitenberg}}, \ and\ \bibinfo {author} {\bibfnamefont
  {J.}~\bibnamefont {Dalibard}},\ }\href {\doibase 10.1038/nphys2378}
  {\bibfield  {journal} {\bibinfo  {journal} {Nature Physics}\ }\textbf
  {\bibinfo {volume} {8}},\ \bibinfo {pages} {645} (\bibinfo {year}
  {2012})}\BibitemShut {NoStop}%
\bibitem [{\citenamefont {Nelson}\ and\ \citenamefont
  {Kosterlitz}(1977)}]{Nelson77a}%
  \BibitemOpen
  \bibfield  {author} {\bibinfo {author} {\bibfnamefont {D.~R.}\ \bibnamefont
  {Nelson}}\ and\ \bibinfo {author} {\bibfnamefont {J.~M.}\ \bibnamefont
  {Kosterlitz}},\ }\href {\doibase 10.1103/PhysRevLett.39.1201} {\bibfield
  {journal} {\bibinfo  {journal} {Phys. Rev. Lett.}\ }\textbf {\bibinfo
  {volume} {39}},\ \bibinfo {pages} {1201} (\bibinfo {year}
  {1977})}\BibitemShut {NoStop}%
\bibitem [{\citenamefont {Minnhagen}\ and\ \citenamefont
  {Warren}(1981)}]{Minnhagen81}%
  \BibitemOpen
  \bibfield  {author} {\bibinfo {author} {\bibfnamefont {P.}~\bibnamefont
  {Minnhagen}}\ and\ \bibinfo {author} {\bibfnamefont {G.~G.}\ \bibnamefont
  {Warren}},\ }\href {\doibase 10.1103/PhysRevB.24.2526} {\bibfield  {journal}
  {\bibinfo  {journal} {Phys. Rev. B}\ }\textbf {\bibinfo {volume} {24}},\
  \bibinfo {pages} {2526} (\bibinfo {year} {1981})}\BibitemShut {NoStop}%
\bibitem [{\citenamefont {Jos\'e}\ \emph {et~al.}(1977)\citenamefont {Jos\'e},
  \citenamefont {Kadanoff}, \citenamefont {Kirkpatrick},\ and\ \citenamefont
  {Nelson}}]{Jose77}%
  \BibitemOpen
  \bibfield  {author} {\bibinfo {author} {\bibfnamefont {J.~V.}\ \bibnamefont
  {Jos\'e}}, \bibinfo {author} {\bibfnamefont {L.~P.}\ \bibnamefont
  {Kadanoff}}, \bibinfo {author} {\bibfnamefont {S.}~\bibnamefont
  {Kirkpatrick}}, \ and\ \bibinfo {author} {\bibfnamefont {D.~R.}\ \bibnamefont
  {Nelson}},\ }\href {\doibase 10.1103/PhysRevB.16.1217} {\bibfield  {journal}
  {\bibinfo  {journal} {Phys. Rev. B}\ }\textbf {\bibinfo {volume} {16}},\
  \bibinfo {pages} {1217} (\bibinfo {year} {1977})}\BibitemShut {NoStop}%
\bibitem [{\citenamefont {Ambegaokar}\ \emph {et~al.}(1980)\citenamefont
  {Ambegaokar}, \citenamefont {Halperin}, \citenamefont {Nelson},\ and\
  \citenamefont {Siggia}}]{Ambegaokar80}%
  \BibitemOpen
  \bibfield  {author} {\bibinfo {author} {\bibfnamefont {V.}~\bibnamefont
  {Ambegaokar}}, \bibinfo {author} {\bibfnamefont {B.~I.}\ \bibnamefont
  {Halperin}}, \bibinfo {author} {\bibfnamefont {D.~R.}\ \bibnamefont
  {Nelson}}, \ and\ \bibinfo {author} {\bibfnamefont {E.~D.}\ \bibnamefont
  {Siggia}},\ }\href {\doibase 10.1103/PhysRevB.21.1806} {\bibfield  {journal}
  {\bibinfo  {journal} {Phys. Rev. B}\ }\textbf {\bibinfo {volume} {21}},\
  \bibinfo {pages} {1806} (\bibinfo {year} {1980})}\BibitemShut {NoStop}%
\bibitem [{\citenamefont {Minnhagen}(1987)}]{Minnhagen87}%
  \BibitemOpen
  \bibfield  {author} {\bibinfo {author} {\bibfnamefont {P.}~\bibnamefont
  {Minnhagen}},\ }\href {\doibase 10.1103/RevModPhys.59.1001} {\bibfield
  {journal} {\bibinfo  {journal} {Rev. Mod. Phys.}\ }\textbf {\bibinfo {volume}
  {59}},\ \bibinfo {pages} {1001} (\bibinfo {year} {1987})}\BibitemShut
  {NoStop}%
\bibitem [{\citenamefont {Gr\"ater}\ and\ \citenamefont
  {Wetterich}(1995)}]{Graeter95}%
  \BibitemOpen
  \bibfield  {author} {\bibinfo {author} {\bibfnamefont {M.}~\bibnamefont
  {Gr\"ater}}\ and\ \bibinfo {author} {\bibfnamefont {C.}~\bibnamefont
  {Wetterich}},\ }\href {\doibase 10.1103/PhysRevLett.75.378} {\bibfield
  {journal} {\bibinfo  {journal} {Phys. Rev. Lett.}\ }\textbf {\bibinfo
  {volume} {75}},\ \bibinfo {pages} {378} (\bibinfo {year} {1995})}\BibitemShut
  {NoStop}%
\bibitem [{\citenamefont {Gersdorff}\ and\ \citenamefont
  {Wetterich}(2001)}]{Gersdorff01}%
  \BibitemOpen
  \bibfield  {author} {\bibinfo {author} {\bibfnamefont {G.~v.}\ \bibnamefont
  {Gersdorff}}\ and\ \bibinfo {author} {\bibfnamefont {C.}~\bibnamefont
  {Wetterich}},\ }\href {\doibase 10.1103/PhysRevB.64.054513} {\bibfield
  {journal} {\bibinfo  {journal} {Phys. Rev. B}\ }\textbf {\bibinfo {volume}
  {64}},\ \bibinfo {pages} {054513} (\bibinfo {year} {2001})}\BibitemShut
  {NoStop}%
\bibitem [{\citenamefont {Nagy}\ \emph {et~al.}(2009)\citenamefont {Nagy},
  \citenamefont {N\'andori}, \citenamefont {Polonyi},\ and\ \citenamefont
  {Sailer}}]{Nagy09}%
  \BibitemOpen
  \bibfield  {author} {\bibinfo {author} {\bibfnamefont {S.}~\bibnamefont
  {Nagy}}, \bibinfo {author} {\bibfnamefont {I.}~\bibnamefont {N\'andori}},
  \bibinfo {author} {\bibfnamefont {J.}~\bibnamefont {Polonyi}}, \ and\
  \bibinfo {author} {\bibfnamefont {K.}~\bibnamefont {Sailer}},\ }\href
  {\doibase 10.1103/PhysRevLett.102.241603} {\bibfield  {journal} {\bibinfo
  {journal} {Phys. Rev. Lett.}\ }\textbf {\bibinfo {volume} {102}},\ \bibinfo
  {pages} {241603} (\bibinfo {year} {2009})}\BibitemShut {NoStop}%
\bibitem [{\citenamefont {Machado}\ and\ \citenamefont
  {Dupuis}(2010)}]{Machado10}%
  \BibitemOpen
  \bibfield  {author} {\bibinfo {author} {\bibfnamefont {T.}~\bibnamefont
  {Machado}}\ and\ \bibinfo {author} {\bibfnamefont {N.}~\bibnamefont
  {Dupuis}},\ }\href {\doibase 10.1103/PhysRevE.82.041128} {\bibfield
  {journal} {\bibinfo  {journal} {Phys. Rev. E}\ }\textbf {\bibinfo {volume}
  {82}},\ \bibinfo {pages} {041128} (\bibinfo {year} {2010})}\BibitemShut
  {NoStop}%
\bibitem [{\citenamefont {Krahl}\ and\ \citenamefont
  {Wetterich}(2007)}]{Krahl07}%
  \BibitemOpen
  \bibfield  {author} {\bibinfo {author} {\bibfnamefont {H.~C.}\ \bibnamefont
  {Krahl}}\ and\ \bibinfo {author} {\bibfnamefont {C.}~\bibnamefont
  {Wetterich}},\ }\href {\doibase
  http://dx.doi.org/10.1016/j.physleta.2007.03.028} {\bibfield  {journal}
  {\bibinfo  {journal} {Phys. Lett. A}\ }\textbf {\bibinfo {volume} {367}},\
  \bibinfo {pages} {263} (\bibinfo {year} {2007})}\BibitemShut {NoStop}%
\bibitem [{\citenamefont {Floerchinger}\ and\ \citenamefont
  {Wetterich}(2009)}]{Floerchinger09a}%
  \BibitemOpen
  \bibfield  {author} {\bibinfo {author} {\bibfnamefont {S.}~\bibnamefont
  {Floerchinger}}\ and\ \bibinfo {author} {\bibfnamefont {C.}~\bibnamefont
  {Wetterich}},\ }\href {\doibase 10.1103/PhysRevA.79.013601} {\bibfield
  {journal} {\bibinfo  {journal} {Phys. Rev. A}\ }\textbf {\bibinfo {volume}
  {79}},\ \bibinfo {eid} {013601} (\bibinfo {year} {2009})}\BibitemShut
  {NoStop}%
\bibitem [{\citenamefont {Ran\c{c}on}\ and\ \citenamefont
  {Dupuis}(2012)}]{Rancon12b}%
  \BibitemOpen
  \bibfield  {author} {\bibinfo {author} {\bibfnamefont {A.}~\bibnamefont
  {Ran\c{c}on}}\ and\ \bibinfo {author} {\bibfnamefont {N.}~\bibnamefont
  {Dupuis}},\ }\href {\doibase 10.1103/PhysRevA.85.063607} {\bibfield
  {journal} {\bibinfo  {journal} {Phys. Rev. A}\ }\textbf {\bibinfo {volume}
  {85}},\ \bibinfo {pages} {063607} (\bibinfo {year} {2012})}\BibitemShut
  {NoStop}%
\bibitem [{\citenamefont {Ran\c{c}on}\ and\ \citenamefont
  {Dupuis}(2013)}]{Rancon13b}%
  \BibitemOpen
  \bibfield  {author} {\bibinfo {author} {\bibfnamefont {A.}~\bibnamefont
  {Ran\c{c}on}}\ and\ \bibinfo {author} {\bibfnamefont {N.}~\bibnamefont
  {Dupuis}},\ }\href {\doibase 10.1209/0295-5075/104/16002} {\bibfield
  {journal} {\bibinfo  {journal} {Europhys. Lett.}\ }\textbf {\bibinfo {volume}
  {104}},\ \bibinfo {pages} {16002} (\bibinfo {year} {2013})}\BibitemShut
  {NoStop}%
\bibitem [{\citenamefont {Ran\c{c}on}(2014)}]{Rancon14a}%
  \BibitemOpen
  \bibfield  {author} {\bibinfo {author} {\bibfnamefont {A.}~\bibnamefont
  {Ran\c{c}on}},\ }\href {\doibase 10.1103/PhysRevB.89.214418} {\bibfield
  {journal} {\bibinfo  {journal} {Phys. Rev. B}\ }\textbf {\bibinfo {volume}
  {89}},\ \bibinfo {pages} {214418} (\bibinfo {year} {2014})}\BibitemShut
  {NoStop}%
\bibitem [{\citenamefont {Prokof'ev}\ \emph {et~al.}(2001)\citenamefont
  {Prokof'ev}, \citenamefont {Ruebenacker},\ and\ \citenamefont
  {Svistunov}}]{Prokofev01}%
  \BibitemOpen
  \bibfield  {author} {\bibinfo {author} {\bibfnamefont {N.}~\bibnamefont
  {Prokof'ev}}, \bibinfo {author} {\bibfnamefont {O.}~\bibnamefont
  {Ruebenacker}}, \ and\ \bibinfo {author} {\bibfnamefont {B.}~\bibnamefont
  {Svistunov}},\ }\href {\doibase 10.1103/PhysRevLett.87.270402} {\bibfield
  {journal} {\bibinfo  {journal} {Phys. Rev. Lett.}\ }\textbf {\bibinfo
  {volume} {87}},\ \bibinfo {pages} {270402} (\bibinfo {year}
  {2001})}\BibitemShut {NoStop}%
\bibitem [{\citenamefont {Prokof'ev}\ and\ \citenamefont
  {Svistunov}(2002)}]{Prokofev02}%
  \BibitemOpen
  \bibfield  {author} {\bibinfo {author} {\bibfnamefont {N.}~\bibnamefont
  {Prokof'ev}}\ and\ \bibinfo {author} {\bibfnamefont {B.}~\bibnamefont
  {Svistunov}},\ }\href {\doibase 10.1103/PhysRevA.66.043608} {\bibfield
  {journal} {\bibinfo  {journal} {Phys. Rev. A}\ }\textbf {\bibinfo {volume}
  {66}},\ \bibinfo {pages} {043608} (\bibinfo {year} {2002})}\BibitemShut
  {NoStop}%
\bibitem [{\citenamefont {Capogrosso-Sansone}\ \emph
  {et~al.}(2008)\citenamefont {Capogrosso-Sansone}, \citenamefont {S\"oyler},
  \citenamefont {Prokof'ev},\ and\ \citenamefont {Svistunov}}]{Capogrosso08}%
  \BibitemOpen
  \bibfield  {author} {\bibinfo {author} {\bibfnamefont {B.}~\bibnamefont
  {Capogrosso-Sansone}}, \bibinfo {author} {\bibfnamefont {S.~G.}\ \bibnamefont
  {S\"oyler}}, \bibinfo {author} {\bibfnamefont {N.}~\bibnamefont {Prokof'ev}},
  \ and\ \bibinfo {author} {\bibfnamefont {B.}~\bibnamefont {Svistunov}},\
  }\href {\doibase 10.1103/PhysRevA.77.015602} {\bibfield  {journal} {\bibinfo
  {journal} {Phys. Rev. A}\ }\textbf {\bibinfo {volume} {77}},\ \bibinfo
  {pages} {015602} (\bibinfo {year} {2008})}\BibitemShut {NoStop}%
\bibitem [{\citenamefont {Wetterich}(1993{\natexlab{a}})}]{Wetterich93a}%
  \BibitemOpen
  \bibfield  {author} {\bibinfo {author} {\bibfnamefont {C.}~\bibnamefont
  {Wetterich}},\ }\href {\doibase 10.1007/BF01474340} {\bibfield  {journal}
  {\bibinfo  {journal} {Z. Phys. C}\ }\textbf {\bibinfo {volume} {57}},\
  \bibinfo {pages} {451} (\bibinfo {year} {1993}{\natexlab{a}})}\BibitemShut
  {NoStop}%
\bibitem [{\citenamefont {Berges}\ \emph {et~al.}(2002)\citenamefont {Berges},
  \citenamefont {Tetradis},\ and\ \citenamefont {Wetterich}}]{Berges02}%
  \BibitemOpen
  \bibfield  {author} {\bibinfo {author} {\bibfnamefont {J.}~\bibnamefont
  {Berges}}, \bibinfo {author} {\bibfnamefont {N.}~\bibnamefont {Tetradis}}, \
  and\ \bibinfo {author} {\bibfnamefont {C.}~\bibnamefont {Wetterich}},\ }\href
  {\doibase doi:10.1016/S0370-1573(01)00098-9} {\bibfield  {journal} {\bibinfo
  {journal} {Phys. Rep.}\ }\textbf {\bibinfo {volume} {363}},\ \bibinfo {pages}
  {223} (\bibinfo {year} {2002})}\BibitemShut {NoStop}%
\bibitem [{\citenamefont {Delamotte}(2012)}]{Delamotte12}%
  \BibitemOpen
  \bibfield  {author} {\bibinfo {author} {\bibfnamefont {B.}~\bibnamefont
  {Delamotte}},\ }in\ \href {\doibase 10.1007/978-3-642-27320-9_2} {\emph
  {\bibinfo {booktitle} {Renormalization Group and Effective Field Theory
  Approaches to Many-Body Systems}}},\ \bibinfo {series} {Lecture Notes in
  Physics}, Vol.\ \bibinfo {volume} {852},\ \bibinfo {editor} {edited by\
  \bibinfo {editor} {\bibfnamefont {A.}~\bibnamefont {Schwenk}}\ and\ \bibinfo
  {editor} {\bibfnamefont {J.}~\bibnamefont {Polonyi}}}\ (\bibinfo  {publisher}
  {Springer Berlin Heidelberg},\ \bibinfo {year} {2012})\ pp.\ \bibinfo {pages}
  {49--132}\BibitemShut {NoStop}%
\bibitem [{\citenamefont {Kopietz}\ \emph {et~al.}(2010)\citenamefont
  {Kopietz}, \citenamefont {Bartosch},\ and\ \citenamefont
  {Sch\"utz}}]{Kopietz_book}%
  \BibitemOpen
  \bibfield  {author} {\bibinfo {author} {\bibfnamefont {P.}~\bibnamefont
  {Kopietz}}, \bibinfo {author} {\bibfnamefont {L.}~\bibnamefont {Bartosch}}, \
  and\ \bibinfo {author} {\bibfnamefont {F.}~\bibnamefont {Sch\"utz}},\ }\href
  {\doibase 10.1007/978-3-642-05094-7} {\emph {\bibinfo {title} {Introduction
  to the Functional Renormalization Group}}}\ (\bibinfo  {publisher}
  {Springer},\ \bibinfo {address} {Berlin},\ \bibinfo {year}
  {2010})\BibitemShut {NoStop}%
\bibitem [{\citenamefont {Wetterich}(1993{\natexlab{b}})}]{Wetterich93}%
  \BibitemOpen
  \bibfield  {author} {\bibinfo {author} {\bibfnamefont {C.}~\bibnamefont
  {Wetterich}},\ }\href {\doibase doi:10.1016/0370-2693(93)90726-X} {\bibfield
  {journal} {\bibinfo  {journal} {Phys. Lett. B}\ }\textbf {\bibinfo {volume}
  {301}},\ \bibinfo {pages} {90} (\bibinfo {year}
  {1993}{\natexlab{b}})}\BibitemShut {NoStop}%
\bibitem [{not({\natexlab{a}})}]{note2}%
  \BibitemOpen
  \href@noop {} \bibinfo {note} {Note that the propagator
  $G_k=(\Gamma^{(2)}_k+R_k)^{-1}$ entering the flow equations includes the
  cutoff function and therefore differs from Eqs.~(\ref{GLT}).}\BibitemShut
  {Stop}%
\bibitem [{\citenamefont {Pogorelov}\ and\ \citenamefont
  {Suslov}(2008)}]{Pogorelov08}%
  \BibitemOpen
  \bibfield  {author} {\bibinfo {author} {\bibfnamefont {A.~A.}\ \bibnamefont
  {Pogorelov}}\ and\ \bibinfo {author} {\bibfnamefont {I.~M.}\ \bibnamefont
  {Suslov}},\ }\href@noop {} {\bibfield  {journal} {\bibinfo  {journal} {Sov.
  Phys. JETP}\ }\textbf {\bibinfo {volume} {106}},\ \bibinfo {pages} {1118}
  (\bibinfo {year} {2008})}\BibitemShut {NoStop}%
\bibitem [{\citenamefont {Campostrini}\ \emph {et~al.}(2006)\citenamefont
  {Campostrini}, \citenamefont {Hasenbusch}, \citenamefont {Pelissetto},\ and\
  \citenamefont {Vicari}}]{Campostrini06}%
  \BibitemOpen
  \bibfield  {author} {\bibinfo {author} {\bibfnamefont {M.}~\bibnamefont
  {Campostrini}}, \bibinfo {author} {\bibfnamefont {M.}~\bibnamefont
  {Hasenbusch}}, \bibinfo {author} {\bibfnamefont {A.}~\bibnamefont
  {Pelissetto}}, \ and\ \bibinfo {author} {\bibfnamefont {E.}~\bibnamefont
  {Vicari}},\ }\href {\doibase 10.1103/PhysRevB.74.144506} {\bibfield
  {journal} {\bibinfo  {journal} {Phys. Rev. B}\ }\textbf {\bibinfo {volume}
  {74}},\ \bibinfo {pages} {144506} (\bibinfo {year} {2006})}\BibitemShut
  {NoStop}%
\bibitem [{\citenamefont {Campostrini}\ \emph {et~al.}(2002)\citenamefont
  {Campostrini}, \citenamefont {Hasenbusch}, \citenamefont {Pelissetto},
  \citenamefont {Rossi},\ and\ \citenamefont {Vicari}}]{Campostrini02}%
  \BibitemOpen
  \bibfield  {author} {\bibinfo {author} {\bibfnamefont {M.}~\bibnamefont
  {Campostrini}}, \bibinfo {author} {\bibfnamefont {M.}~\bibnamefont
  {Hasenbusch}}, \bibinfo {author} {\bibfnamefont {A.}~\bibnamefont
  {Pelissetto}}, \bibinfo {author} {\bibfnamefont {P.}~\bibnamefont {Rossi}}, \
  and\ \bibinfo {author} {\bibfnamefont {E.}~\bibnamefont {Vicari}},\ }\href
  {\doibase 10.1103/PhysRevB.65.144520} {\bibfield  {journal} {\bibinfo
  {journal} {Phys. Rev. B}\ }\textbf {\bibinfo {volume} {65}},\ \bibinfo
  {pages} {144520} (\bibinfo {year} {2002})}\BibitemShut {NoStop}%
\bibitem [{\citenamefont {Canet}\ \emph {et~al.}(2003)\citenamefont {Canet},
  \citenamefont {Delamotte}, \citenamefont {Mouhanna},\ and\ \citenamefont
  {Vidal}}]{Canet03b}%
  \BibitemOpen
  \bibfield  {author} {\bibinfo {author} {\bibfnamefont {L.}~\bibnamefont
  {Canet}}, \bibinfo {author} {\bibfnamefont {B.}~\bibnamefont {Delamotte}},
  \bibinfo {author} {\bibfnamefont {D.}~\bibnamefont {Mouhanna}}, \ and\
  \bibinfo {author} {\bibfnamefont {J.}~\bibnamefont {Vidal}},\ }\href
  {\doibase 10.1103/PhysRevB.68.064421} {\bibfield  {journal} {\bibinfo
  {journal} {Phys. Rev. B}\ }\textbf {\bibinfo {volume} {68}},\ \bibinfo
  {pages} {064421} (\bibinfo {year} {2003})}\BibitemShut {NoStop}%
\bibitem [{not({\natexlab{b}})}]{note6}%
  \BibitemOpen
  \href@noop {} \bibinfo {note} {For a detailed
  discussion of this issue, see L. Canet, Hugues Chat\'e, and B. Delamotte
  (unpublished).}\BibitemShut {Stop}%
\bibitem [{not({\natexlab{c}})}]{note1}%
  \BibitemOpen
  \href@noop {} \bibinfo {note} {If we take a microscopic
  cutoff $\Lambda$ of the order of $0.1~\mathring{A}^{-1}$,
  $k=\Lambda\,e^{-20}$ then corresponds to a length scale of the order of
  0.5~m.}\BibitemShut {Stop}%
\bibitem [{not({\natexlab{d}})}]{note5}%
  \BibitemOpen
  \href@noop {} \bibinfo {note} {A nonpositive propagator
  means that the Legendre transform of the free energy $-\ln Z[\J]$ is not
  convex.}\BibitemShut {Stop}%
\bibitem [{not({\natexlab{e}})}]{note7}%
  \BibitemOpen
  \href@noop {} \bibinfo {note} {This additional
  approximation is not mentioned in Ref.~\cite{Gersdorff01}, but is described
  in detail in Ref.~\cite{Gersdorff00}.}\BibitemShut {Stop}%
\bibitem [{not({\natexlab{f}})}]{note3}%
  \BibitemOpen
  \href@noop {} \bibinfo {note} {The characteristic
  length scale $k_c^{-1}$ is analog to the healing length of a
  superfluid.}\BibitemShut {Stop}%
\bibitem [{not({\natexlab{g}})}]{note4}%
  \BibitemOpen
  \href@noop {} \bibinfo {note} {The long-distance
  behavior of the propagator can be obtained from the effective action
  $S[\theta] = (\rho_s(T)/2) \int d^2r\, (\nablabf\theta)^2$, where the phase
  variable $\theta$ is free of vortex singularities and represents the spinwave
  degrees of freedom. An elementary calculation of the propagator $G(\r-\r')
  \sim \mean{e^{i\theta(\r)-i\theta(\r')}}$ then yields an anomalous dimension
  $\eta(T)=1/2\pi\rho_s(T)$.}\BibitemShut {Stop}%
\bibitem [{not({\natexlab{h}})}]{note9}%
  \BibitemOpen
  \href@noop {} \bibinfo {note} {The universality of
  the product $bc=\pi/2$ follows from the result of Appendix~\ref{app_bkt_flow}
  noting that $r_0-r_{0c}\propto T-\Tkt$.}\BibitemShut {Stop}%
\bibitem [{\citenamefont {{Benfatto}}\ \emph {et~al.}(2013)\citenamefont
  {{Benfatto}}, \citenamefont {{Castellani}},\ and\ \citenamefont
  {{Giamarchi}}}]{Benfatto13}%
  \BibitemOpen
  \bibfield  {author} {\bibinfo {author} {\bibfnamefont {L.}~\bibnamefont
  {{Benfatto}}}, \bibinfo {author} {\bibfnamefont {C.}~\bibnamefont
  {{Castellani}}}, \ and\ \bibinfo {author} {\bibfnamefont {T.}~\bibnamefont
  {{Giamarchi}}},\ } in\ \href {\doibase 10.1142/9789814417648_0005} {\emph {\bibinfo
  {booktitle} {40 Years of Berezinskii-Kosterlitz-Thouless Theory}}},\ \bibinfo {editor} {edited by\
  \bibinfo {editor} {\bibfnamefont {J.~V.}\ \bibnamefont {{Jos{\'e}}}}}\
  (\bibinfo  {publisher} {World Scientific, Singapore},\ \bibinfo {year}
  {2013})\ pp.\ \bibinfo {pages} {161--199}\BibitemShut {NoStop}%
\bibitem [{\citenamefont {Strack}\ \emph {et~al.}(2008)\citenamefont {Strack},
  \citenamefont {Gersch},\ and\ \citenamefont {Metzner}}]{Strack08}%
  \BibitemOpen
  \bibfield  {author} {\bibinfo {author} {\bibfnamefont {P.}~\bibnamefont
  {Strack}}, \bibinfo {author} {\bibfnamefont {R.}~\bibnamefont {Gersch}}, \
  and\ \bibinfo {author} {\bibfnamefont {W.}~\bibnamefont {Metzner}},\ }\href
  {\doibase 10.1103/PhysRevB.78.014522} {\bibfield  {journal} {\bibinfo
  {journal} {Phys. Rev. B}\ }\textbf {\bibinfo {volume} {78}},\ \bibinfo
  {pages} {014522} (\bibinfo {year} {2008})}\BibitemShut {NoStop}%
\bibitem [{\citenamefont {Obert}\ \emph {et~al.}(2013)\citenamefont {Obert},
  \citenamefont {Husemann},\ and\ \citenamefont {Metzner}}]{Obert13}%
  \BibitemOpen
  \bibfield  {author} {\bibinfo {author} {\bibfnamefont {B.}~\bibnamefont
  {Obert}}, \bibinfo {author} {\bibfnamefont {C.}~\bibnamefont {Husemann}}, \
  and\ \bibinfo {author} {\bibfnamefont {W.}~\bibnamefont {Metzner}},\ }\href
  {\doibase 10.1103/PhysRevB.88.144508} {\bibfield  {journal} {\bibinfo
  {journal} {Phys. Rev. B}\ }\textbf {\bibinfo {volume} {88}},\ \bibinfo
  {pages} {144508} (\bibinfo {year} {2013})}\BibitemShut {NoStop}%
\bibitem [{\citenamefont {Strack}\ and\ \citenamefont
  {Jakubczyk}(2014)}]{Strack14}%
  \BibitemOpen
  \bibfield  {author} {\bibinfo {author} {\bibfnamefont {P.}~\bibnamefont
  {Strack}}\ and\ \bibinfo {author} {\bibfnamefont {P.}~\bibnamefont
  {Jakubczyk}},\ }\href {\doibase 10.1103/PhysRevX.4.021012} {\bibfield
  {journal} {\bibinfo  {journal} {Phys. Rev. X}\ }\textbf {\bibinfo {volume}
  {4}},\ \bibinfo {pages} {021012} (\bibinfo {year} {2014})}\BibitemShut
  {NoStop}%
\bibitem [{\citenamefont {Eberlein}\ and\ \citenamefont
  {Metzner}(2013)}]{Eberlein13a}%
  \BibitemOpen
  \bibfield  {author} {\bibinfo {author} {\bibfnamefont {A.}~\bibnamefont
  {Eberlein}}\ and\ \bibinfo {author} {\bibfnamefont {W.}~\bibnamefont
  {Metzner}},\ }\href {\doibase 10.1103/PhysRevB.87.174523} {\bibfield
  {journal} {\bibinfo  {journal} {Phys. Rev. B}\ }\textbf {\bibinfo {volume}
  {87}},\ \bibinfo {pages} {174523} (\bibinfo {year} {2013})}\BibitemShut
  {NoStop}%
\bibitem [{\citenamefont {Eberlein}\ and\ \citenamefont
  {Metzner}(2014)}]{Eberlein14}%
  \BibitemOpen
  \bibfield  {author} {\bibinfo {author} {\bibfnamefont {A.}~\bibnamefont
  {Eberlein}}\ and\ \bibinfo {author} {\bibfnamefont {W.}~\bibnamefont
  {Metzner}},\ }\href {\doibase 10.1103/PhysRevB.89.035126} {\bibfield
  {journal} {\bibinfo  {journal} {Phys. Rev. B}\ }\textbf {\bibinfo {volume}
  {89}},\ \bibinfo {pages} {035126} (\bibinfo {year} {2014})}\BibitemShut
  {NoStop}%
\bibitem [{\citenamefont {Friederich}\ \emph {et~al.}(2010)\citenamefont
  {Friederich}, \citenamefont {Krahl},\ and\ \citenamefont
  {Wetterich}}]{Friederich10}%
  \BibitemOpen
  \bibfield  {author} {\bibinfo {author} {\bibfnamefont {S.}~\bibnamefont
  {Friederich}}, \bibinfo {author} {\bibfnamefont {H.~C.}\ \bibnamefont
  {Krahl}}, \ and\ \bibinfo {author} {\bibfnamefont {C.}~\bibnamefont
  {Wetterich}},\ }\href {\doibase 10.1103/PhysRevB.81.235108} {\bibfield
  {journal} {\bibinfo  {journal} {Phys. Rev. B}\ }\textbf {\bibinfo {volume}
  {81}},\ \bibinfo {pages} {235108} (\bibinfo {year} {2010})}\BibitemShut
  {NoStop}%
\bibitem [{\citenamefont {Friederich}\ \emph {et~al.}(2011)\citenamefont
  {Friederich}, \citenamefont {Krahl},\ and\ \citenamefont
  {Wetterich}}]{Friederich11}%
  \BibitemOpen
  \bibfield  {author} {\bibinfo {author} {\bibfnamefont {S.}~\bibnamefont
  {Friederich}}, \bibinfo {author} {\bibfnamefont {H.~C.}\ \bibnamefont
  {Krahl}}, \ and\ \bibinfo {author} {\bibfnamefont {C.}~\bibnamefont
  {Wetterich}},\ }\href {\doibase 10.1103/PhysRevB.83.155125} {\bibfield
  {journal} {\bibinfo  {journal} {Phys. Rev. B}\ }\textbf {\bibinfo {volume}
  {83}},\ \bibinfo {pages} {155125} (\bibinfo {year} {2011})}\BibitemShut
  {NoStop}%
\bibitem [{\citenamefont {v.~Gersdorff}(2000)}]{Gersdorff00}%
  \BibitemOpen
  \bibfield  {author} {\bibinfo {author} {\bibfnamefont {G.}~\bibnamefont
  {v.~Gersdorff}},\ }\href@noop {} {\enquote {\bibinfo {title}
  {Zweidimensionale o($n$)-symmetrische systeme im formalismus der exakten
  renormierungsgruppe},}\ } (\bibinfo {year} {2000}),\ \bibinfo {note}
  {diplomarbeit, Heidelberg University (unpublished).}\BibitemShut {Stop}%
\bibitem [{\citenamefont {Chaikin}\ and\ \citenamefont
  {Lubensky}(1995)}]{Chaikin_book}%
  \BibitemOpen
  \bibfield  {author} {\bibinfo {author} {\bibfnamefont {P.~M.}\ \bibnamefont
  {Chaikin}}\ and\ \bibinfo {author} {\bibfnamefont {T.~C.}\ \bibnamefont
  {Lubensky}},\ }\href@noop {} {\emph {\bibinfo {title} {Principles of
  Condensed Matter Physics}}}\ (\bibinfo  {publisher} {Cambridge University
  Press},\ \bibinfo {year} {1995})\BibitemShut {NoStop}%
\end{thebibliography}


%

\end{document}